\documentclass[twoside,twocolumn,english]{IEEEtran}
\usepackage[T1]{fontenc}
\usepackage[latin9]{inputenc}
\usepackage{float}
\usepackage{fancybox}
\usepackage{calc}
\usepackage{textcomp}
\usepackage{amsmath}
\usepackage{amssymb}
\usepackage{graphicx}
\usepackage{setspace}
\usepackage{esint}
\makeatletter

\floatstyle{ruled}
\newfloat{algorithm}{tbp}{loa}
\providecommand{\algorithmname}{Algorithm}
\floatname{algorithm}{\protect\algorithmname}

\twocolumn

\makeatother

\usepackage{babel}
\begin{document}
\global\long\def\bd{{\bf d}}

\global\long\def\real{\mathbb{R}}

\global\long\def\bx{{\bf x}}

\global\long\def\ba{\mathbf{a}}

\global\long\def\bb{{\bf b}}

\global\long\def\bg{{\bf g}}

\global\long\def\bd{{\bf d}}

\global\long\def\bo{{\bf 0}}

\global\long\def\by{{\bf y}}
 \global\long\def\bz{{\bf z}}
 \global\long\def\bh{{\bf h}}

\global\long\def\bw{{\bf w}}

\global\long\def\bba{{\bf A}}
 \global\long\def\bbb{{\bf B}}

\global\long\def\bbf{{\bf F}}
 \global\long\def\bbx{{\bf X}}
 \global\long\def\bbu{{\bf U}}
 \global\long\def\bbi{{\bf I}}
 \global\long\def\argmin{{\displaystyle \mathop{\mbox{argmin}}}}
 \global\long\def\argmax{{\displaystyle \mathop{\mbox{argmax}}}}

\title{Phase Retrieval with Application to Optical Imaging}

\author{\IEEEauthorblockN{Yoav Shechtman\IEEEauthorrefmark{1}\IEEEauthorrefmark{2},
Yonina C. Eldar\IEEEauthorrefmark{3}, Oren Cohen\IEEEauthorrefmark{1},
Henry N. Chapman\IEEEauthorrefmark{4}\IEEEauthorrefmark{5}\IEEEauthorrefmark{6},
Jianwei Miao\IEEEauthorrefmark{7} and Mordechai Segev\IEEEauthorrefmark{1}\\
} \IEEEauthorblockA{\IEEEauthorrefmark{1}Department of Physics
Technion, Israel Institute of Technology, Israel 32000}\\
\IEEEauthorblockA{\IEEEauthorrefmark{2}Department of Chemistry,
Stanford University, Stanford, California 94305, USA}\\
 \IEEEauthorblockA{\IEEEauthorrefmark{3}Department of Electrical
Engineering, Technion, Israel Institute of Technology, Israel 32000}\\
\IEEEauthorblockA{\IEEEauthorrefmark{4}Center for Free-Electron
Laser Science, DESY, Notkestrasse 85, 22607 Hamburg, Germany }\\
 \IEEEauthorblockA{\IEEEauthorrefmark{5}University of Hamburg,
Luruper Chaussee 149, 22761 Hamburg, Germany }\\
\IEEEauthorblockA{\IEEEauthorrefmark{6}Center for Ultrafast Imaging,
22607 Hamburg, Germany }\\
 \IEEEauthorblockA{\IEEEauthorrefmark{7}Department of Physics and
Astronomy, and California NanoSystems Institute, University of California,
Los Angeles, CA 90095, USA \\
Email: yoavsh@stanford.edu}}

\maketitle

\section{Introduction}

The problem of phase retrieval, namely \textendash{} the recovery
of a function given the magnitude of its Fourier transform - arises
in various fields of science and engineering, including electron microscopy,
crystallography, astronomy, and optical imaging. Exploring phase retrieval
in optical settings, specifically when the light originates from a
laser, is natural, because optical detection devices (e.g., ccd cameras,
photosensitive films, the human eye) cannot measure the phase of a
light wave. This is because, generally, optical measurement devices
that rely on converting photons to electrons (current) do not allow
direct recording of the phase: the electromagnetic field oscillates
at rates $\sim10^{15}$ Hz, which no electronic measurement devices
can follow. Indeed, optical measurement / detection systems measure
the photon flux, which is proportional to the magnitude squared of
the field, not the phase. Consequently, measuring the phase of optical
waves (electromagnetic fields oscillating at $10^{15}$ Hz and higher)
involves additional complexity, typically by interfering it with another
(known) field, in the process of holography.

Interestingly, electromagnetic fields do have some other features
which make them amenable for algorithmic phase retrieval: their far-field
corresponds to the Fourier transform of their near-field. More specifically,
given a \textquotedblleft{}mask\textquotedblright{} that superimposes
some structure (an image) on a quasi-monochromatic coherent field
at some plane in space, the electromagnetic field structure at a large
enough distance from that plane is given by the Fourier transform
of the image multiplied by a known quadratic phase factor. Thus, measuring
the \emph{far-field}, magnitude and phase, would facilitate recovery
of the optical image (the wavefield). However, as noted above, the
optical phase cannot be measured directly by an electronic detector.
Here is where algorithmic phase retrieval comes into play, offering
a means for recovering the phase given the measurement of the magnitude
of the optical far-field and some prior knowledge. 

The purpose of this review article is to provide a contemporary review
of phase retrieval in optical imaging. It begins with historical background
section that also explains the physical setting, followed by a section
on the mathematical formulation of the problem. The fourth section
discusses existing algorithms, while the fifth section describes various
contemporary applications. The last section discusses additional physical
settings where algorithmic phase retrieval is important, identifies
current challenges and provides a long term vision. This review article
provides a contemporary overview of phase retrieval in optical imaging,
linking the relevant optical physics to the signal processing methods
and algorithms. Our goal is to describe the current state of the art
in this area, identify challenges, and suggest vision and areas where
signal processing methods can have a large impact on optical imaging
and on the world of imaging at large, with applications in a variety
of fields ranging from biology and chemistry to physics and engineering.

\section{Historical Background}

\begin{figure}
\begin{centering}
\includegraphics[width=0.9\columnwidth]{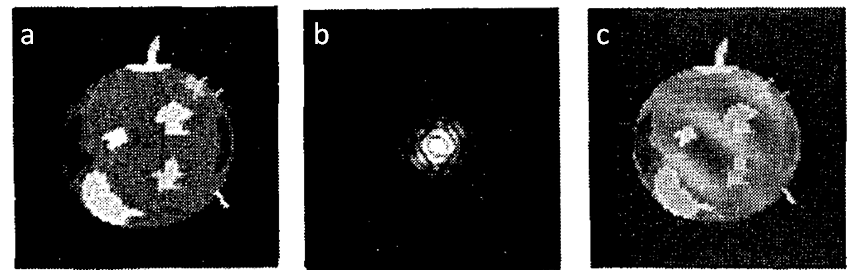}
\par\end{centering}

\caption{\label{fig:Numerical-2D-phase}Numerical 2D phase retrieval example,
adapted from Fienup's 1978 paper \cite{fienup_reconstruction_1978}.
(a) Test object. (b) Fourier magnitude (c) Reconstruction results
(using HIO - see Fig. \ref{fig:GS_HIO}b for details)}

\end{figure}

Algorithmic phase retrieval offers an alternative means for recovering
the phase structure of optical images, without requiring sophisticated
measuring setups as in holography. These approaches typically rely
on some advanced information in order to facilitate recovery. Back
in 1952, Sayre envisioned, in the context of crystallography, that
the phase information of a scattered wave may be recovered if the
intensity pattern at and between the Bragg peaks of the diffracted
wave is finely measured \cite{sayre_implications_1952}. In crystallography,
the material structure under study is periodic (a crystal), hence
the far-field information contains naturally strong peaks reflecting
the Fourier transform of the periodic information. Measuring the fine
features in the Fourier transform enabled the recovery of the phase
in some simple cases. Twenty six years later, in 1978, Fienup developed
algorithms for retrieving phases of 2D images from their Fourier modulus
and constraints such as non-negativity, and a known support of the
image \cite{fienup_reconstruction_1978} (See Fig. \ref{fig:Numerical-2D-phase}).
In the early eighties, the idea of phase retrieval created a flurry
of follow up work, partly because those times signified great hope
for realizing an optical computer, of which phase retrieval was supposed
to be a key ingredient. However, in the 1980s and 1990s, with the
understanding that an optical computer is unrealistic, the interest
in algorithmic phase retrieval diminished. Towards the end of the
millennium, optical phase retrieval started to come back into contemporary
optics research, with the interest arising from a completely different
direction: the community of researchers experimenting with X-ray imaging,
where new X-ray sources (undulators and synchrotrons) were developed.
The wide-spread interest of this field was mainly generated by the
first experimental recording and reconstruction of a continuous diffraction
pattern (Fourier magnitude squared) of a non-crystalline (non-periodic)
test object by Miao and collaborators in 1999 \cite{miao_extending_1999}.

The reasons for the revival of optical phase retrieval, in 1999, were
actually quite subtle. One goal of optical imaging systems is to increase
resolution, that is, to image smaller and smaller features. But, as
known since Ernst Abbe's work in 1873, the highest attainable resolution
in diffraction imaging (so-called the \emph{diffraction limit}) is
comparable to the wavelength of the light. For visible light, this
diffraction limit corresponds to fraction of microns. Consequently,
features on the molecular scale cannot be viewed with visible light
in a microscope. One could argue then, why not simply use electromagnetic
waves of a much shorter wavelength, say, in the hard X-ray regime,
where the wavelength is comparable to atomic resolution? The reason
is that lens-like devices and other optical components in this spectral
region suffer from very large aberrations and are very difficult to
make due to fact that refractive indices of materials in this wavelength
regime are close to one. On the other hand, algorithmic phase retrieval
is of course not limited by the quality of lenses; however it requires
very low noise detectors. 

An additional problem is that as resolution is improved (that is,
as voxel elements in the recovered image are smaller in size), the
number of photons per unit area must obviously increase to provide
a reasonable SNR. This means that the required exposure time to obtain
a given signal level must increase as $(1/d)^{4}$, with $d$ being
the resolution length, assumed to be larger than atomic scales \cite{Howells20094}.
This, in turn, creates another problem: X-ray photons are highly energetic.
The atomic cross section for photoabsorption is usually much higher
than for elastic scattering, meaning that for every photon that contributes
to the diffraction pattern (the measured Fourier magnitude), a considerable
greater number of photons are absorbed by the sample. This energy
dissipates in the sample first by photoionisation and the breakage
of bonds, followed by a cascade of collisional ionisation by free
electrons and, at longer timescales, a destruction of the sample due
to radiolysis, heating, or even ablation of the sample. Such radiation
damage hinders the ability to recover the structure of molecules:
the measured far-field intensity (Fourier magnitude) also reflects
the structural damages, rather than providing information about the
true molecular structure \cite{sayre1995x}. A solution to this problem
was suggested by Solem and Chapline in the 1980's. They proposed to
record images (or holograms in their case) with pulses that are shorter
than the timescale for the X-ray damage to manifest itself at a particular
resolution. They predicted that picosecond pulses would be required
to image at nanometer length scales \cite{solem_microholography_1982}.
Towards the late nineties, with the growing promise in constructing
X-ray lasers that generate ultrashort pulses on the femtosecond scale,
it was suggested that such pulses could even outrun damage processes
at atomic length scales\cite{neutze_potential_2000}. However, forming
a direct image in this way would still require high quality optical
components (lenses, mirrors) in the X-ray regime, which do not currently
exist. This is because creating lenses for the hard X-ray wavelength
regime requires fabrication at picometer resolution, much smaller
even than the Bohr radius of atoms. Likewise, while mirrors for X-rays
do exist, their best resolution is on the scale of many nanometers,
much larger than the features one would want to resolve in imaging
of molecules, for example. 

\begin{figure*}
\centering{}%
\Ovalbox{\begin{minipage}[t]{1\textwidth}%
\begin{center}
\textbf{Coherent Diffractive Imaging (CDI)}
\par\end{center}

In the basic CDI setup (forward scattering), an object is illuminated
by a quasi-monochromatic coherent wave, and the diffracted intensity
is measured (Fig. \ref{fig:CDI-setup:-A}). When the object is small
and the intensity is measured far away, the measured intensity is
proportional to the magnitude of the Fourier transform of the wave
at the object plane, with appropriate spatial scaling. 

\begin{center}
\includegraphics[width=0.6\columnwidth]{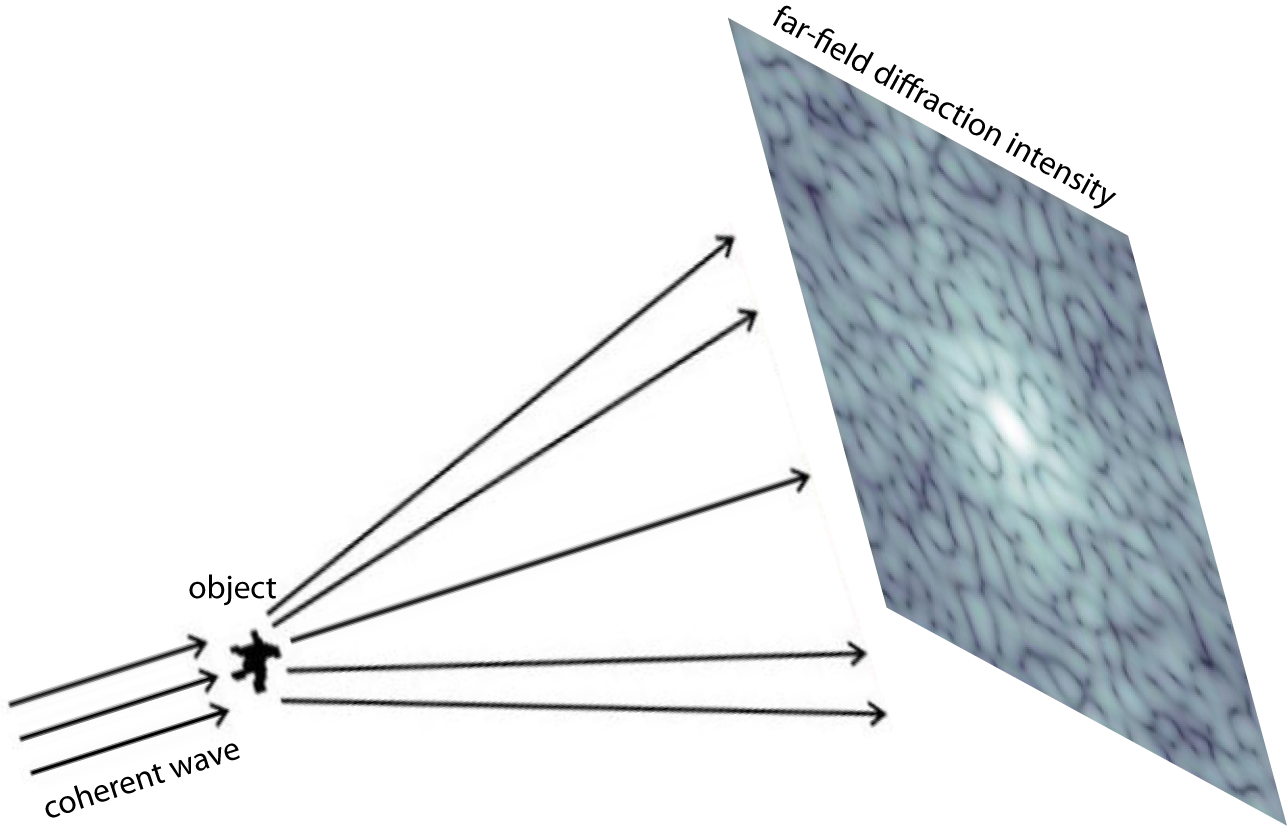}
\par\end{center}

\caption{\label{fig:CDI-setup:-A}A forward-scattering CDI setup: A coherent
wave diffracts from an object (the \emph{sought information}), and
produces a far-field intensity pattern corresponding to the magnitude
of the Fourier transform of the object. }

In optics terms, when the Fresnel number is small ($N_{F}=\frac{a^{2}}{\lambda d}<<1$,
where $a$ is a radius confining the object in the object plane, $d$
is the distance between the object and the measured intensity plane,
and $\lambda$ is the wavelength of the light), the relation between
the measured intensity $I_{out}$ and the wave at the object plane
$E_{in}$, is given by \cite{saleh_fundamentals_2007}: 
\[
I_{out}(x,y)\propto\bigg|\hat{E}_{in}\bigg(\frac{x}{\lambda d},\frac{y}{\lambda d}\bigg)\bigg|^{2}
\]

with $\hat{E}_{in}=\mathcal{F}\{E_{in}\}$, and $\mathcal{F}$ denoting
the Fourier transform. Once the far field intensity is measured, the
goal is to recover $E_{in}$ (which is equivalent to recovering the
object) from $I_{out}$. This requires solving the phase retrieval
problem, which is attempted using an algorithm such as the ones described
in this review paper.%
\end{minipage}}
\end{figure*}

The difficulties outlined above in direct X-ray imaging leave no choice
but to use alternative methods to recover the structure of nanometric
samples. Here is where phase retrieval can make its highest impact.
Placing an area detector far enough from the sample to record the
far-field diffraction intensity (which is approximately proportional
to the squared magnitude of the Fourier transform of the image, if
the coherence length of the X-ray wave is larger than the sample size
\cite{miao2002high,spence2004coherence}), together with appropriate
constraints on the support of the sample, enable the recovery of the
image at nanometric resolution. Indeed, the phase information has
been shown numerically and experimentally to be retrieved in this
fashion in various examples \cite{fienup_reconstruction_1978,elser2003solution,marchesini2007invited,bauschke2003hybrid,luke2005relaxed,rodriguez2013oversampling}.
The combination of X-ray diffraction, oversampling and phase retrieval
has launched the currently very active field called \emph{Coherent
Diffraction Imaging} or CDI \cite{miao_extending_1999}. In CDI, an
object is illuminated with a coherent wave, and the far field diffraction
intensity pattern (corresponding to the Fourier magnitude of the object)
is measured. The problem then is to recover the object from the measured
far-field intensity (See box on Coherent Diffractive Imaging and Fig.
\ref{fig:CDI-setup:-A} within). Since its first experimental demonstration,
CDI has been applied to image a wide range of samples using synchrotron
radiation \cite{miao2002high,robinson2001reconstruction,williams2003three,miao2003imaging,chapman2006high,abbey2008keyhole,dierolf_ptychographic_2010,nam2013imaging},
X-ray free electron lasers (XFELs) \cite{barty_ultrafast_2008-1,mancuso2009coherent,chapman_femtosecond_2006,seibert_single_2011,loh_fractal_2012},
high harmonic generation \cite{sandberg_lensless_2007,ravasio2009single,chen2009multiple,seaberg_ultrahigh_2011},
soft X-ray laser \cite{sandberg2008high}, optical laser \cite{bertolotti2012non},
and electrons \cite{miao2002atomic,zuo2003atomic,putkunz2012atom}.
Several readable reviews on the development and implementation of
phase-retrieval algorithms for the specific application of CDI were
written by Marchesini \cite{marchesini2007invited}, Thibault and
Elser \cite{thibault2010x} and Nugent \cite{nugent2010coherent}.
Presently, one of the most challenging problems in CDI is towards
3D structural determination of large protein molecules \cite{neutze_potential_2000,miao2001approach}.
There has been ongoing progress towards this goal during the past
decade. In 2006, Chapman et al., demonstrated the CDI of a test sample
using intense ultra-short single pulse from free electron laser, relying
on recording a diffraction pattern before the sample was destroyed
\cite{chapman_femtosecond_2006}. Recently, the technique was implemented
for high-resolution imaging of isolated sub-micron objects, such as
herpesvirus \cite{song_quantitative_2008}, mimivirus \cite{seibert_single_2011}
and aerosol particles such as soot \cite{loh_fractal_2012}. 

From a theoretical and algorithm perspective, phase retrieval is a
difficult problem, in many cases lacking a unique solution. Furthermore,
even with the existence of a unique solution, there is not necessarily
a guarantee that it can be found algorithmically. Nevertheless, as
reasoned above, phase retrieval algorithms and applications have benefited
from a surge of research in recent years, in large part due to various
new imaging techniques in optics. This trend has begun impacting the
signal processing community as well \textendash{} the past few years
have witnessed growing interest within this community in developing
new approaches to phase retrieval by using tools of modern optimization
theory \cite{candes_phaselift:_2012,candes2013phaseSIAM}. More recent
work has begun exploring connections between phase retrieval and structure-based
information processing \cite{balan2006signal,moravec2007compressive,fannjiang_absolute_2012,bandeira_saving_2013,beck_sparsity_2012,szameit_sparsity-based_2012}.
For example, it has been shown that, by exploiting the sparsity of
many optical images, one can develop powerful phase retrieval methods
that allow for increased resolution considerably beyond Abbe's diffraction
limit, resolving features smaller than $1/5$ of the wavelength \cite{szameit_sparsity-based_2012}.
The relationship between the fields of sparsity and optical imaging
has led to an important generalization of the basic principles of
sparsity-based reconstruction to nonlinear measurement systems \cite{moravec2007compressive,shechtman_sparsity_2011-3,beck_sparsity_2012,ohlsson_compressive_2011,bahmani_greedy_2011,jaganathan_recovery_2012,eldar_phase_2012,shechtman_gespar:_2013,shechtman_efficient_2013,Shechtman:13}.
Here too, optics played an important role in signal processing: since
the phase retrieval problem is inherently mathematically nonlinear
(i.e., the sought signal is related to the measurements nonlinearly),
employing sparsity-based concepts in phase retrieval required genuine
modifications to the linear sparsity-based algorithms known from the
field of compressed sensing \cite{eldar_compressed_2012}. We believe
that this field will grow steadily in the next few years, with rapid
development of coherent X-ray sources worldwide \cite{emma2010first,ishikawa2012compact}
and more researchers contributing to the theory, algorithms and practice
of nonlinear sparse recovery.

\section{Mathematical Formulation}

\subsection{Problem Formulation}

Consider the discretized 1D real space distribution function of an
object: $\mathbf{x}\in\mathbb{C}^{N}$ (extension of the formulation
to higher dimensions is trivial). In CDI, for example, this corresponds
to the transmittance function of the object. The fact that $\mathbf{x}$
is in general complex, corresponds physically to the fact that the
electromagnetic field emanating from different points on the object
has not only magnitude but also phase (as is always the case, for
example, when 3D objects are illuminated and light is reflected from
point at different planes). The 1D discrete Fourier transform (DFT)
of $\mathbf{x}$ is given by:
\begin{equation}
X[k]=\sum_{n=0}^{N-1}x[n]e^{-j2\pi\frac{kn}{N}},\,\,\,\,\, k=0,1,...,N-1.
\end{equation}
The term \emph{oversampled DFT} used in this paper will refer to an
$M$ point DFT of $\bx\in\mathbb{C}^{N}$ with $M>N$:
\begin{equation}
X[k]=\sum_{n=0}^{N-1}x[n]e^{-j2\pi\frac{kn}{M}},\,\,\,\,\, k=0,1,\dots,M-1.
\end{equation}

Recovery of $\mathbf{x}$ from measurement of $\bbx$ can be achieved
by simply applying the inverse-DFT operator. Writing $X[k]=|X[k]|\cdot e^{j\phi[k]}$,
the Fourier phase retrieval problem is to recover \textbf{$\bx$}
when only the magnitude of $X$ is measured, i.e. to recover $x[n]$
given $|X[k]|$. Since the DFT operator is bijective, this is equivalent
to recovering the phase of $X[k]$, namely, $\phi[k]$ - hence the
name \emph{phase retrieval}. Denote by $\mathbf{\hat{x}}$ the vector
$\mathbf{x}$ after padding with $N-1$ zeros. The autocorrelation
sequence of $\mathbf{\hat{x}}$ is then defined as: 
\begin{equation}
g[m]=\sum_{i=\max\{1,m+1\}}^{N}\hat{x}_{i}\overline{\hat{x}{}_{i-m}},\,\,\,\,\, m=-(N-1),\ldots,N-1.
\end{equation}
It is well known that the DFT of $g[m]$, denoted by $G[k]$, satisfies
$G[k]=|X[k]|^{2}$. Thus, the problem of recovering a signal from
its Fourier magnitude is equivalent to the problem of recovering a
signal from the autocorrelation sequence of its oversampled version.

Continuous phase retrieval can be defined similarly to its discrete
counterpart, as the recovery of a 1D signal $f(x)$ from its continuous
Fourier magnitude:
\[
|F(\nu)|=\Big|\int_{\mathbb{R}}f(x)\exp(-j2\pi vx)dx\Big|.
\]
The actual objects of interest, electromagnetic fields, are usually
described by continuous functions. However, since the data acquisition
is digitized (by CCD camera and alike), and the processing is done
digitally, we shall mostly treat here the discrete case. 

The Fourier phase retrieval problem is as a special case of the more
general phase retrieval problem, where we are given measurements:
\begin{equation}
y_{k}=|\langle\ba_{k},\bx\rangle|^{2},\,\,\, k=1,\dots,M,\label{eq:4}
\end{equation}
with $\ba_{k}$ denoting the measurement vectors. In discrete 1D Fourier
phase retrieval the measurement vectors are given by $\ba_{k}[n]=e^{-j2\pi\frac{kn}{M}}$.
For mathematical analysis, it is often easier to treat the case where
the measurements are random (i.e. $\ba_{k}$ are random vectors),
as this allows uniqueness guarantees that are otherwise hard to obtain
\cite{waldspurger_phase_2012,candes_phaselift:_2012,eldar_phase_2012,netrapalli2013phase,li2012sparse}.
Nevertheless, more structured measurements have also been investigated
\cite{gross2013partial}. 

\begin{figure*}
\begin{centering}
\includegraphics[width=0.8\textwidth]{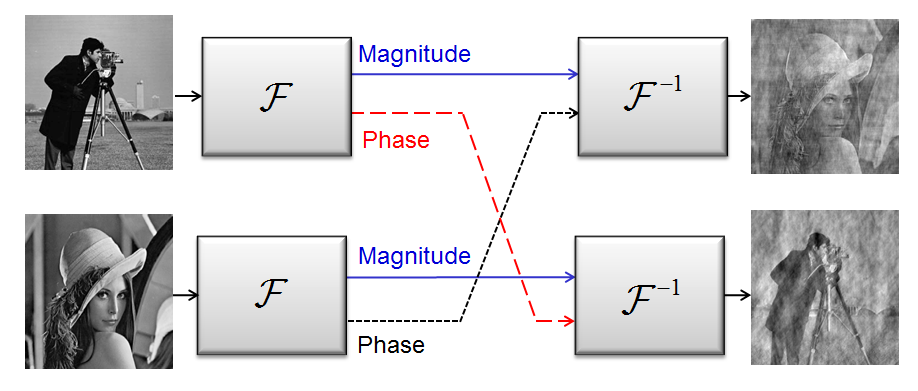}
\par\end{centering}

\caption{\label{fig:The-importance-of}The importance of Fourier phase. Two
images, Cameraman and Lenna, are Fourier transformed. After swapping
their phases, they are inverse Fourier transformed. The result clearly
demonstrates the importance of phase information for image recovery.}

\end{figure*}

Before proceeding to the mathematical methodology, it is important
to highlight the significance of knowing the Fourier phase. In fact,
it is well known that knowledge of the Fourier phase is crucial in
recovering an object from its Fourier transform \cite{oppenheim1981importance}.
Many times the Fourier phase contains more information than the Fourier
magnitude, as can be seen in the synthetic example shown in Fig. \ref{fig:The-importance-of}.
The figure shows the result of the following numerical experiment:
Two images (Cameraman and Lenna) are Fourier transformed. The phases
of their transforms are swapped, and subsequently they are inverse
Fourier transformed. It is evident, for this quite arbitrary example,
that the Fourier phase contains a significant amount of information
about the images. In crystallography, this phenomenon is the source
of genuine concern of \emph{phase bias} of molecular models (such
as used in molecular replacement) in refined structures.\textbf{ }

In the remainder of this section we discuss uniqueness of the phase
retrieval problem, i.e. under what conditions is the solution to the
phase problem unique? It is worth noting that, while the discussion
of theoretical uniqueness guarantees is important and interesting,
the lack of such guarantees does not prevent practical applications
from producing excellent reconstruction results in many settings.

\subsection{Uniqueness }

\subsubsection{Fourier measurements}

The recovery of a signal from its Fourier magnitude alone, in general,
does not yield a unique solution. This section will review the main
existing theoretical results regarding phase-retrieval uniqueness. 

First, there are so called \emph{trivial} ambiguities that are always
present. The following three transformations (or any combination of
them) conserve Fourier magnitude: 

1. Global phase shift: $x[n]\Rightarrow x[n]\cdot e^{j\phi_{0}}$

2. Conjugate inversion: $x[n]\Rightarrow\overline{x[-n]}$

3. Spatial shift: $x[n]\Rightarrow x[n+n_{0}]$.

Second, there are non-trivial ambiguities, the situation of which
varies for different problem-dimensions. In the 1D problem there is
no uniqueness \textendash{} i.e. there are multiple 1D signals with
the same Fourier magnitude. Even if the support of the signal is bounded
within a known range, uniqueness does not exist \cite{hofstetter_construction_1964}.
Any pair of 1D signals having the same autocorrelation function yields
the same Fourier magnitude, as the two are connected by a Fourier
transform. Consider for example the two vectors $\mathbf{u}=[1\,\,0\,\,-2\,\,0\,\,-2]^{T}$
and $\mathbf{v=}[(1-\sqrt{3})\,\,0\,\,1\,\,0\,\,(1+\sqrt{3})]^{T}$.
Both of these vectors have the same support, and yield the same autocorrelation
function $g[m]=[-2,0,2,0,9,0,2,0,-2]$. Therefore, they are indistinguishable
by their Fourier magnitude, even though they are not \emph{trivially}
equivalent.

For higher dimensions (2D and above), Bruck and Sodin \cite{bruck1979ambiguity},
Hayes \cite{hayes_reconstruction_1982}, and Bates \cite{bates_fourier_1982}
have shown that, with the exception of a set of signals of measure
zero, a real $d\geq2$ dimensional signal with support $\mathbf{N}=[N_{1}\dots N_{d}]$,
namely $x(n_{1},\dots,n_{d})=0$ whenever $n_{k}<0$ or $n_{k}\geq N_{k}$
for $k=1,\dots,d$ is uniquely specified by the magnitude of its continuous
Fourier transform, up to the trivial ambiguities mentioned above.
Furthermore, the magnitude of the oversampled $\mathbf{M}$ point
DFT sequence of the signal, with $\mathbf{M\geq}2\mathbf{N}-1$ (where
the inequality holds in every dimension), is sufficient to guarantee
uniqueness. The problematic set of signals that are not uniquely defined
by their Fourier magnitudes are those having a reducible Z transform:
denoting the $d$ dimensional $Z$ transform of $x$ by $X(z_{1},\dots,z_{d})=\sum_{n_{1}}\cdots\sum_{n_{d}}x(n_{1},\dots,n_{d})z_{1}^{-n_{1}}\cdots z_{d}^{-n_{d}}$
, $X(\mathbf{z})$ is said to be reducible if it can be written as
$X(\mathbf{z})=X_{1}(\mathbf{z})X_{2}(\mathbf{z})$, where $X_{1}(\mathbf{z})$
and $X_{2}(\mathbf{z})$ are both polynomials in $\mathbf{z}$ with
degree $p>0$. It is important to note that in practice, for typical
images, a number of samples smaller than $2\mathbf{N}-1$ is many
times sufficient (even $\mathbf{M=N}$ can work \cite{miao_phase_1998}),
however the exact guarantees relating the number of samples to the
type of images remains an open question.

Additional prior information about the sought signal, other than its
support, can be incorporated, and will naturally improve the conditioning
of the problem. For example, knowledge of the Fourier phase sign (i.e.
1 bit of phase information) has been shown \cite{van1983signal} to
yield uniqueness with some restrictions on the signal (specifically
that the signal is real and its Z transform has no zeros on the unit
circle). A different, popular type of prior knowledge that has been
used recently in various applications \cite{eldar_compressed_2012,elad_sparse_2010},
is that the signal $\mathbf{x}\in\mathbb{C}^{N}$ is sparse - i.e.
contains only a small number $k$ of nonzero elements, with $k\ll N$.
The exact locations and values of the nonzero elements are not known
a-priori. In this case, it has been shown \cite{ranieri2013phase}
that knowledge of the full autocorrelation sequence of a 1D $k$ sparse
real signal $\bx$ is sufficient in order to uniquely define $\bx$,
as long as $k\neq6$ and the autocorrelation sequence is \emph{collision
free}. A vector $\bx$ is said to have a collision free autocorrelation
sequence if $x(i)-x(j)\neq x(k)-x(l)$, for all distinct $i,j,k,l\in\{1,\dots N\}$
that are the locations of distinct nonzero values in $\bx$. In addition,
under these conditions, only $M$ Fourier magnitude measurements are
sufficient to uniquely define the autocorrelation sequence and therefore
the signal $\bx$, as long as $M$ is prime and $M\geq k^{2}-k+1$
\cite{ohlsson2013conditions}. An interesting perspective relating
phase retrieval to the Turnpike problem, namely, reconstructing a
set of integers from their pairwise distances, is presented in \cite{jaganathan_sparse_2013}.Using
this approach, the authors prove uniqueness with high probability,
for random signals having a non-periodic support.

\subsubsection{General measurements}

Considering inner products with general, non-Fourier (typically -
random) measurement vectors, allows simpler derivation of theoretical
guarantees. There have been several theoretical results relating the
number and the nature of the measurements that are required for uniqueness,
mostly dealing with random measurement vectors. The work of Balan
\cite{balan2006signal} implies that for real signals in $\mathbb{R}^{N}$,
$2N-1$ random measurements are needed, provided that they are full-spark,
i.e. that every subset of $N$ measurement vectors spans $\mathbb{R}^{N}$
\cite{bandeira_saving_2013}. This result was later extended to the
complex case \cite{bandeira_saving_2013}, where it is conjectured
that $4N-4$ \emph{generic} measurements, as defined in \cite{bandeira_saving_2013},
are sufficient for bijectivity. In terms of stability, i.e. when the
measurements are noisy, it has been proven \cite{eldar_phase_2012}
that on the order of $N\log(N)$ measurements (or $k\log(N)$ measurements
in the $k$ sparse case) are sufficient for stable uniqueness. It
was also shown that minimizing the (nonconvex) least-squares objective:
$\sum|y_{i}^{2}-|\langle\ba_{i},\bx\rangle|^{2}|^{p}$, with $1<p\leq2$,
yields the correct solution under these conditions \cite{eldar_phase_2012}.
For the noiseless case, any $k$-sparse vector in $\mathbb{R}^{N}$
has been shown to be uniquely determined by $4k-1$ random Gaussian
intensity measurements with high probability \cite{ohlsson2013conditions}.

To study the injectivity of general (i.e. not necessarily random)
measurements, the \emph{complement-property} has been introduced in
\cite{balan2006signal} for the real case. An extension was presented
in \cite{bandeira_saving_2013} for the complex setting. A set of
measurement vectors $\{\ba_{i}\}_{i=1}^{M}$ with $a_{i}\in\mathbb{R}^{N}$
satisfies the complement property if for every $S\subseteq\{1,\dots,M\}$,
either $ $$\{\ba_{i}\}_{i\in S}$ or $ $$\{\ba_{i}\}_{i\in S^{C}}$
span $\mathbb{R}^{N}$. It has been shown in \cite{balan2006signal}
that the mapping constructed by $y_{i}=|\langle\ba_{i},\bx\rangle|,\,\,\, i=1,\dots,N$
is injective if and only if the measurement set satisfies the complement
property. This poses a lower limit on the number of necessary measurements
$M>2N-1$.

The results reviewed in this section are summarized in Table \ref{tab:Phase-retrieval--}.
In addition, there is a large amount of work on phase retrieval uniqueness
under different conditions, e.g. when the phase is known only approximately
\cite{osherovich2011approximate}, or from redundant masked Fourier
measurements \cite{candes2013phase,fannjiang_absolute_2012}.

\begin{table*}
\centering{}\includegraphics[width=0.8\textwidth]{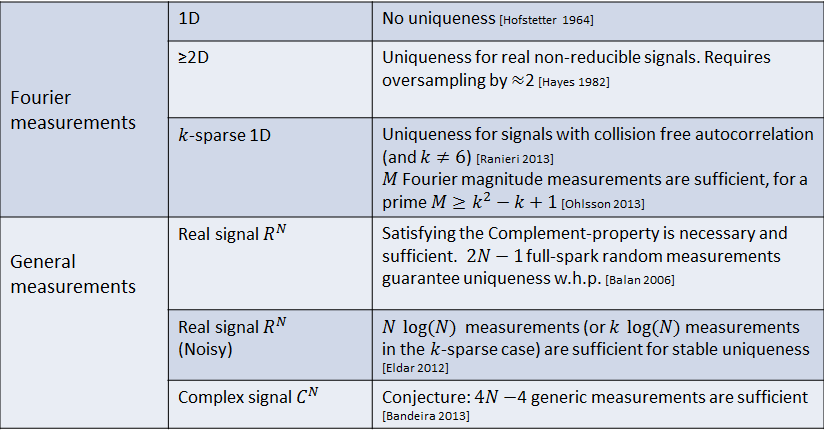}\caption{\label{tab:Phase-retrieval--}Phase retrieval - uniqueness}
\end{table*}

\section{Algorithms}

Despite the uniqueness guarantees, no known general solution method
exists to actually find the unknown signal from its Fourier magnitude
given the other constraints. Over the years, several approaches have
been suggested for solving the phase retrieval problem, with the popular
ones being alternating projection algorithms \cite{gerchberg1972practical,fienup_reconstruction_1978,fienup_phase_1982}.
In addition, methods were suggested attempting to solve phase retrieval
problems by using exposures with different masks \cite{candes2013phase},
or images obtained at different propagation planes \cite{Streibl19846,soto_improved_2007}.
Another method to obtain additional information is scanning CDI, (also
termed ptychography) \cite{hoppe_beugung_1969,rodenburg_ptychography_2008,dierolf_ptychographic_2010},
which uses several different illumination patterns to obtain coherent
diffraction images. 

In this section we survey existing phase retrieval algorithms, including
general algorithms (sub-Section \ref{sub:General-algorithms}), and
sparsity based algorithms, i.e. algorithms exploiting prior knowledge
in the form of signal-sparsity (sub-Section \ref{sub:Sparsity-based-algorithms}).

\subsection{\label{sub:General-algorithms}General algorithms}

The general phase retrieval problem we wish to solve can be formulated
as the following least squares problem, or empirical risk minimization:
\begin{equation}
\min_{\bx}\sum_{k=1}^{M}\big(y_{k}-|\langle\ba_{k},\bx\rangle|^{2}\big)^{2},\label{eq:LS}
\end{equation}
with $\mathbf{y}$ being the measurements and $\ba_{k}$ being the
measurement vectors defined in (\ref{eq:4}). In general we can replace
the square in the objective by any power $p$. Unfortunately, this
is a non-convex problem, and it is not clear how to find a global
minimum even if one exists. In this section we describe several approaches
that have been suggested to deal with this problem, and types of prior
information that can be incorporated into these methods in order to
increase the probability of convergence to the true solution.

\subsubsection{\label{sub:Alternating-projections}Alternating projections}

\begin{figure*}
\begin{centering}
\includegraphics[width=1\textwidth]{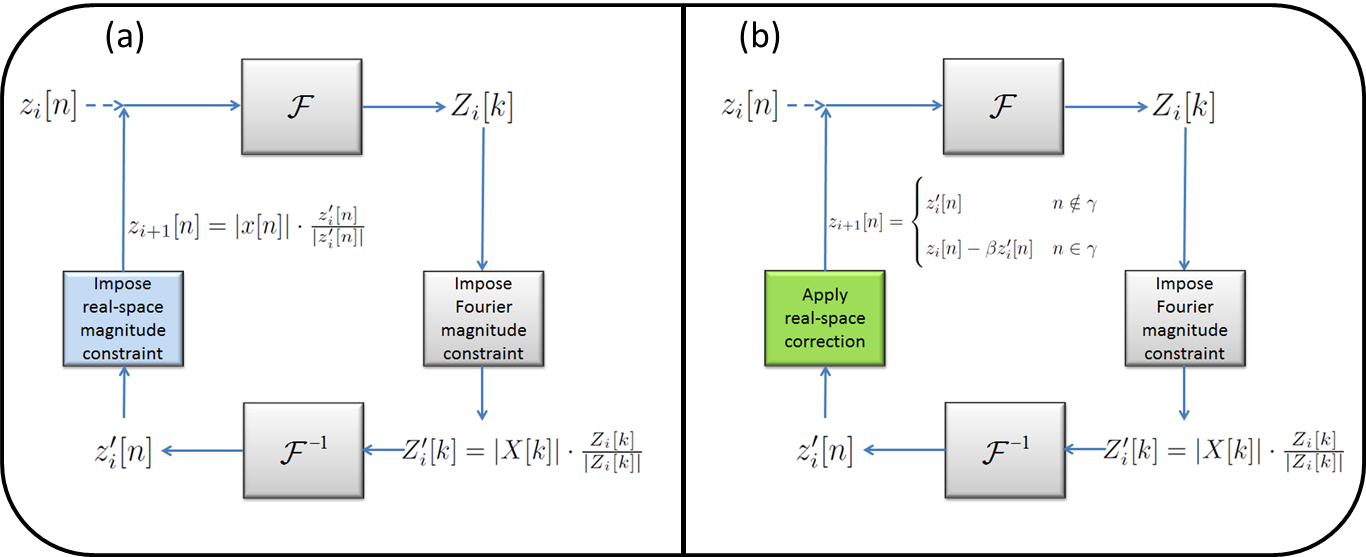}
\par\end{centering}

\caption{\label{fig:GS_HIO}(a) Block diagram of the Gerchberg-Saxton algorithm.
(b) Block diagram of the Fienup HIO algorithm. The algorithms differ
in their fourth (colored) step.}

\end{figure*}

The most popular class of phase retrieval methods is based on alternate
projections. These methods were pioneered by the work of Gerchberg
and Saxton \cite{gerchberg1972practical}, dealing with the closely
related problem of recovering a complex image from magnitude measurements
at two different planes - the real (\emph{imaging}) plane and Fourier
(\emph{diffraction}) plane. The original Gerchberg-Saxton (GS) algorithm
consists of iteratively imposing the real-plane and Fourier-plane
constraints, namely, the measured real-space magnitude $|x[n]|$ and
Fourier magnitude $|X[k]|$, as illustrated in Fig. \ref{fig:GS_HIO}a.
The GS algorithm is described in Algorithm \ref{alg:GS}. The recovery
error, defined as $E_{i}=\sum_{k}\Big||Z_{i}[k]|-|X[k]|\Big|^{2}$
is easily shown to be monotonically decreasing with $i$ \cite{fienup_phase_1982}.
Despite this fact, recovery to the true solution is not guaranteed,
as the algorithm can converge to a local minimum.

\begin{algorithm}
\caption{\label{alg:GS}Gerchberg-Saxton (GS)}

\begin{singlespace}
\textbf{Input:} $|x[n]|,|X[k]|,\epsilon$
\end{singlespace}

$|x[n]|$ - Real space magnitude

$|X[k]|$ - Fourier magnitude

$\epsilon$ - Error threshold

\textbf{Output:} $z[n]$ - a vector that conforms with both magnitude
constraints, i.e.: $|z[n]|=|x[n]|$, and $|Z[k]|=|X[k]|$, where $Z[k]$
is the DFT of $z[n]$

\begin{singlespace}
\line(1,0){250}

\textbf{Initialization}. Choose initial $z_{0}[n]=|x[n]|\exp(\phi[n])$
(e.g. with a random $\phi[n]$).

\textbf{General Step ($i=1,2,\ldots$):}
\end{singlespace}
\begin{enumerate}
\item Fourier transform $z_{i}[n]$ to obtain $Z_{i}[k]$
\item Keep current Fourier phase, but impose Fourier magnitude constraint:
$Z_{i}'[k]=|X[k]|\cdot\frac{Z_{i}[k]}{|Z_{i}[k]|}$
\item Inverse Fourier transform $Z_{i}'[k]$ to obtain $z_{i}'[n]$
\item Keep current real-space phase, but impose real-space magnitude constraint:
$z_{i+1}[n]=|x[n]|\cdot\frac{z_{i}'[n]}{|z_{i}'[n]|}$
\item Go to 1
\end{enumerate}
\begin{singlespace}
\textbf{Until} $E_{i}=\sum_{k}\Big||Z_{i}[k]|-|X[k]|\Big|^{2}\leq\epsilon$\end{singlespace}
\end{algorithm}

Extending the GS projection ideas further, Fienup in 1978 \cite{fienup_reconstruction_1978}
suggested a modified version of the GS algorithm, in which the real-space
magnitude constraints may be replaced by other types of constraints,
in addition to consistency with the measured Fourier magnitude. The
real-space constraints may be for example non-negativity, a known
signal support, namely $x[i]=0$ for all $i>N_{0}$, where $N_{0}$
is known (or approximately known), or both. The basic framework of
the Fienup methods is similar to GS - in fact, the first three steps
are identical. Step 4), however, replaces imposing the real-space
magnitude constraint by applying a correction to the real-space estimate.
Some possible variants to this step were also suggested \cite{fienup_phase_1982}.
Here we describe the one most commonly used, referred to as the hybrid-input
output (HIO) method, which consists of the following correction step:
\begin{enumerate}
\item [4)]\setcounter{enumi}{4}Obtain $z_{i+1}[n]$ by applying a correction
to the real-space image estimate: 
\begin{equation}
z_{i+1}[n]=\begin{cases}
z_{i}'[n], & n\notin\gamma\\
z_{i}[n]-\beta z_{i}'[n], & n\in\gamma,
\end{cases}
\end{equation}
 with $\beta$ being a small parameter, and $\gamma$ being the set
of indices for which $z_{i}'[n]$ violates the real-space constraints.
\end{enumerate}
The real-space constraint violation may be a support violation (signal
is nonzero where it should be zero), or a non-negativity violation. 

The Fienup algorithm is represented schematically in Fig. \ref{fig:GS_HIO}b.
There is no proof that the HIO algorithm converges. It is also known
to be sensitive to the accuracy of the prior information (e.g. the
real-space support needs to be tightly known, especially in the complex
signal case \cite{Fienup:87}). Nonetheless, in practice, the simple
HIO based methods are commonly used in optical phase retrieval applications
such as CDI \cite{miao2005quantitative,miao2006three,jiang2010quantitative}.
Other variants of the correction step include the Input-Output method,
and the Output-Output method \cite{fienup_phase_1982}, corresponding
respectively to 

\begin{equation}
\begin{array}{c}
z_{i+1}[n]=\begin{cases}
z_{i}[n], & n\notin\gamma\\
z_{i}[n]-\beta z_{i}'[n], & n\in\gamma,
\end{cases}\\
z_{i+1}[n]=\begin{cases}
z_{i}'[n], & n\notin\gamma\\
z_{i}'[n]-\beta z_{i}'[n], & n\in\gamma.
\end{cases}
\end{array}
\end{equation}

An important feature of the HIO algorithm is its empirical ability
to avoid local minima and converge to a global minimum for noise-free
oversampled diffraction patterns. However, when there is high noise
present in the diffraction intensity, HIO suffers from several limitations.
First, the algorithm sometimes becomes stagnant and fails to converge
to a global minimum. Second, a support has to be pre-defined. Third,
the image oscillates as a function of the iteration. Over the years,
various algorithms have been developed to overcome these limitations,
including the combination of HIO and the error-reduction (ER) algorithm
\cite{fienup_phase_1982}, difference map \cite{elser2003solution},
hybrid projection reflection \cite{bauschke2003hybrid}, guided hybrid
input-output (GHIO) \cite{chen2007application}, relaxed averaged
alternating reflectors (RAAR) \cite{luke2005relaxed}, noise robust
(NR)-HIO \cite{Martin:12}, and oversampling smoothness (OSS) \cite{rodriguez2013oversampling}.
An analysis of iterative phase retrieval algorithms from a convex
optimization perspective can be found in \cite{bauschke_phase_2002}.

As an exapmle, the recently proposed OSS algorithm exhibits improved
performance over HIO and its variants in many settings. OSS is based
on Fienup iterations, with an added smoothing Gaussian filter applied
to the off-support region in the real space object in each iteration.
Namely, the fourth step in HIO is replaced by:

\begin{equation}
\begin{array}{c}
z''[n]=\begin{cases}
z_{i}'[n], & n\notin\gamma\\
z_{i}[n]-\beta z_{i}'[n], & n\in\gamma,
\end{cases}\\
z_{i+1}[n]=\begin{cases}
z_{i}''[n], & n\notin\gamma\\
\mathcal{F}\{Z_{i}''[k]W[k]\}, & n\in\gamma,
\end{cases}
\end{array}
\end{equation}
where $W[k]$ is a Gaussian function, with its variance decreasing
with iterations. A quantitative comparison for a specific example
between OSS and HIO can be found in Section \ref{sub:Quantitative-comparisons-of}.
For a comparison and numerical investigation of several alternate
projection algorithms see for example \cite{marchesini2007invited,rodriguez2013oversampling}. 

As performance of the Fienup methods is dependent on the initial points,
it is possible and recommended to try several initializations. In
\cite{netrapalli2013phase}, the authors consider a clever method
for initial point selection, and show that for the random Gaussian
measurement case, the resulting iterations yield a solution arbitrarily
close to the true vector.

\subsubsection{Semi-Definite Programming (SDP) based algorithms}

An alternative recently developed to solve the phase retrieval problem
is based on semidefinite relaxation \cite{vandenberghe1996semidefinite,shechtman_sparsity_2011-3,candes2013phaseSIAM}.
The method relies on the observation that (\ref{eq:4}) describes
a set of quadratic equations, which can be re-written as linear equations
in a higher dimension. Specifically, define the $N\times N$ matrix
$\bbx=\bx\bx^{*}$. The measurements (\ref{eq:4}) are then linear
in $\bbx$:

\begin{equation}
y_{k}=|\langle\mathbf{a}_{k},\bx\rangle|^{2}=\bx^{*}\ba_{k}\ba_{k}^{*}\bx=\bx^{*}\bba_{k}\bx=\textnormal{Tr}(\bba_{k}\bbx),\label{eq:trace eq}
\end{equation}
where $\bba_{k}=\ba_{k}\ba_{k}^{*}$. Our problem is then to find
a matrix $\bbx=\bx\bx^{*}$ that satisfies (\ref{eq:trace eq}). The
constraint $\bbx=\bx\bx^{*}$ is equivalent to the requirement that
$\bbx$ has rank one, and is positive semi-definite, which we denote
by $\bbx\succeq0$ . Therefore, finding a vector $\bx$ satisfying
(\ref{eq:4}) can be formulated as:

\begin{equation}
\begin{array}{ll}
\textnormal{find} & \mathbf{X}\\
\mbox{s.t.} & y_{k}=\textnormal{Tr}(\bba_{k}\bbx),\,\,\, k=1,\dots,M,\\
 & \bbx\succeq0,\\
 & \textnormal{rank}(\bbx)=1.
\end{array}\label{eq:trace optimization}
\end{equation}

Problem (\ref{eq:trace optimization}) is equivalent to the following
rank minimization problem:

\begin{equation}
\begin{array}{ll}
\textnormal{\ensuremath{\min}} & \textnormal{rank}(\bbx)\\
\mbox{s.t.} & y_{k}=\textnormal{Tr}(\bba_{k}\bbx),\,\,\, k=1,\dots,M,\\
 & \bbx\succeq0.
\end{array}\label{eq:7}
\end{equation}
Unfortunately, rank minimization is a hard combinatorial problem.
However, since the constraints in (\ref{eq:7}) are convex (in fact
linear), one might try to relax the minimum rank objective, for example
by replacing it with minimization of $\textnormal{Tr}(\bbx)$. This
approach is referred to as PhaseLift \cite{candes2013phaseSIAM}.
Alternatively, one may use the log-det reweighted rank minimization
heuristic suggested in \cite{fazel2003log}, which is the approach
followed in \cite{shechtman_sparsity_2011-3,candes_phaselift:_2012}.
In \cite{candes_phaselift:_2012} it is shown that PhaseLift yields
the true vector $\bx$ with large probability, when the measurements
are random Gaussian and $M\sim O(N\log N)$.

The SDP approach requires matrix lifting, namely, replacing the sought
vector with a higher dimensional matrix, followed by solving a higher
dimensional problem. It is therefore, in principle, more computationally
demanding than the alternating projection approaches, or greedy methods,
which will be discussed in the next section. In addition, in general
there is neither a guarantee that the rank minimization process will
yield a rank-one matrix, nor that the true solution is found, even
if there is a unique solution.

\subsection{\label{sub:Sparsity-based-algorithms}Sparsity based algorithms}

A specific kind of prior knowledge that can be incorporated into the
phase retrieval problem to help regularize it, is the fact that the
sought real-space object is sparse in some known representation (See
sparse linear problems box). This means that the object $\mathbf{x}$
can be written as:

\begin{equation}
\bx=\mathbf{\Psi\boldsymbol{\alpha}}
\end{equation}
with $\mathbf{\boldsymbol{\Psi}}$ being a representation matrix (the
sparsity basis), and $\boldsymbol{\alpha}$ being a sparse vector,
i.e. a vector containing a small number of nonzero coefficients. The
simplest example is when the object is composed of a small number
of point sources (in which case $\boldsymbol{\Psi}$ is the identity
matrix). Armed with such prior knowledge, one can hope to improve
the performance of phase retrieval algorithms, by limiting the search
for the true vector to the set of sparse vectors. There are several
different ways that such knowledge can be incorporated, which are
described in this subsection.

\subsubsection*{Alternating projections with sparsity prior}

The Fienup algorithm, described in Section \ref{sub:Alternating-projections},
allows in principle incorporation of various types of general knowledge
about the object, including sparsity \cite{moravec2007compressive,mukherjee_iterative_2012}.
The method in \cite{mukherjee_iterative_2012}, for example, is based
on the Fienup iterations, with the first three steps remaining unchanged.
Step 4, is replaced by projection and thresholding. Assuming an invertible
$\boldsymbol{\Psi}$ and a $k$ sparse vector $\boldsymbol{\alpha}$
such that $\bx=\boldsymbol{\Psi\alpha}$: 
\begin{enumerate}
\item [4)]\setcounter{enumi}{4}Obtain $z_{i+1}[n]$ by projecting $z_{i}'[n]$
onto $\boldsymbol{\Psi}$, thresholding, and projecting back. Namely:

\begin{enumerate}
\item Calculate $\boldsymbol{\alpha}{}_{i}=\boldsymbol{\Psi}^{-1}\mathbf{z}_{i}'$
\item Keep only the $k$ largest elements of $\mathbf{|a}_{i}|$, setting
the rest to zero.
\item Set $\mathbf{z}_{i+1}=\boldsymbol{\Psi}\boldsymbol{\alpha}_{i}$.
\end{enumerate}
\end{enumerate}
Similarly to the GS method, the error here can be shown to be nonincreasing,
so that convergence to a local minimum is guaranteed \cite{mukherjee_iterative_2012}.

Note, that while this method is suggested in \cite{mukherjee_iterative_2012}
for an orthonormal basis $\boldsymbol{\Psi}$, it can be easily modified
to accommodate a non-invertible $\boldsymbol{\Psi}$. This can be
done by replacing parts (a)+(b) with finding a sparse solution $\ba_{i}$
to $\mathbf{z}_{i}=\boldsymbol{\Psi}\ba_{i}$, using any sparse solution
method \cite{eldar_compressed_2012}. 

\begin{figure*}
\Ovalbox{\begin{minipage}[t]{1\textwidth}%
\begin{singlespace}
\begin{center}
\textbf{Sparse Linear Problems}
\par\end{center}
\end{singlespace}

\begin{singlespace}
Finding sparse solutions to sets of equations is a topic that has
drawn much attention in recent years \cite{candes_robust_2006,donoho_compressed_2006,eldar_compressed_2012,elad_sparse_2010}.
Consider the linear system:
\begin{equation}
\by=\bba\bx\label{eq:sparse_lin}
\end{equation}

with $\by$ being a set of $M$ linear measurements, $\bba$ being
an $M\times N$ measurement matrix, and $\bx$ being the unknown length-$N$
vector. When the system is underdetermined (i.e. $M<N$), there are
infinitely many possible solutions $\mathbf{x}$. A key result of
the theory of sparse recovery is that adding the constraint that $\bx$
is sparse, i.e. contains only a few nonzero entries guarantees a unique
solution to (\ref{eq:sparse_lin}), under general conditions for $\bba$.
One such condition is based on the \emph{coherence} of $\bba$ \cite{donoho_optimally_2003}:
\begin{equation}
||\bx||_{0}\leq\frac{1}{2}\bigg(1+\frac{1}{\mu}\bigg)\label{eq:coherence}
\end{equation}

with $||\bx||_{0}$ being the number of nonzero entries in $\bx$,
and the coherence defined by: 
\begin{equation}
\mu=\max_{i,j}\frac{<\bba_{i},\bba_{j}>}{||\bba_{i}||\cdot||\bba_{j}||}.\label{eq:coherenceDef}
\end{equation}
Here, we denote by $\bba_{i}$ the $i$th column of $\bba$, and by
$||\bba_{i}||$ its Euclidean norm. 
\end{singlespace}

Under (\ref{eq:coherence}), one can find the unique solution to (\ref{eq:sparse_lin})
by solving 
\begin{equation}
\min_{\bx}||\bx||_{0}\,\,\,\textnormal{s.t.}\,\,\,\by=\bba\bx.\label{eq:l0}
\end{equation}

~~~~Unfortunately, (\ref{eq:l0}) is an NP-hard combinatorial
problem. However, many methods have been develop to approximately
solve (\ref{eq:l0}). One class of such methods consists of greedy
algorithms such as Orthogonal Matching Pursuit \cite{pati1993orthogonal}.
Another popular method is based on convex relaxation of the $l_{0}$
norm to an $l_{1}$ norm \cite{chen_atomic_1998}, which yields the
convex problem:
\begin{equation}
\min_{\bx}||\bx||_{1}\,\,\,\textnormal{s.t.\,\,\,}\by=\bba\bx.\label{eq:l1}
\end{equation}

\begin{singlespace}
In fact, under the condition (\ref{eq:coherence}), it has been shown
\cite{donoho_optimally_2003} that the solution to (\ref{eq:l1})
is equal to the solution of (\ref{eq:l0}).
\end{singlespace}

~~~~Another important criterion to evaluate the recovery ability
in sparse linear problems of the form (\ref{eq:sparse_lin}) is the
restricted isometry property (RIP) \cite{candes_decoding_2005} of
$\bba$, defined as follows: For an an $M\times N$ matrix $\bba$
(with $M<N$), define $\delta_{k}$ as the smallest value such that
for every submatrix $\bba_{k}$ composed of $k$ columns of $\bba$,
it holds that 
\begin{equation}
(1-\delta_{k})||\mathbf{x}||_{2}^{2}\leq||\bba_{k}\bx||_{2}^{2}\leq(1+\delta_{k})||\mathbf{x}||_{2}^{2},\,\,\,\,\,\,\,\forall\mathbf{x}\in\mathbb{R}^{k}.\,\,\,\,\,\label{eq:RIP}
\end{equation}

The RIP is therefore a measure of whether $\bba$ preserves the energy
of any $k$ sparse signal - which is the case if $\delta_{k}$ is
small. In the context of sparse recovery, it is used to prove uniqueness
and noise-robustness results. For example, if $\bba$ is such that
$\delta_{2k}<\sqrt{2}-1$, then solving (\ref{eq:l1}) will yield
the unique sparse solution to (\ref{eq:sparse_lin}). In practice,
it is combinatorially difficult to calculate the RIP of a given matrix.
However, certain random matrices can be shown to have 'good RIP' with
high probability. For example, an $M\times N$ iid Gaussian matrix
obeys the $k$-RIP with high probability, for $M\sim O(k\log(N/k))$
\cite{candes_robust_2006}. This is one of the reasons that random
matrices are favorable for sparse sensing. %
\end{minipage}}

\end{figure*}

\subsubsection*{SDP based methods with sparsity prior}

SDP based methods can also be modified to account for prior knowledge
of signal sparsity. The incorporation of sparsity can be performed
in several different ways. The first work to suggest sparsity-based
SDP phase-retrieval came from the domain of optics, and dealt with
partially spatially-incoherent illumination \cite{shechtman_sparsity_2011-3}.
This work actually considered a theoretical problem of greater complexity,
combining phase retrieval with sub-wavelength imaging. Experimental
results on sub-wavelength CDI can be found in \cite{szameit_sparsity-based_2012},
where the sought signal is an actual optical image with subwavelength
features, and the measured data corresponds to the Fourier magnitude
sampled by a camera at the focal plane of a microscope lens. 

\begin{figure*}
\begin{centering}
\includegraphics[width=0.7\textwidth]{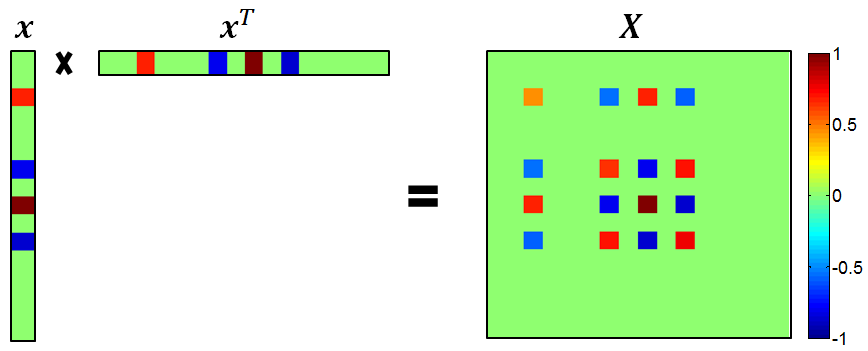}\caption{\label{fig:Sparse-vector-outer}Sparse vector outer product yields
a sparse matrix}

\par\end{centering}

\end{figure*}

The method suggested in \cite{shechtman_sparsity_2011-3}, dubbed
QCS for Quadratic Compressed Sensing, is based on adding sparsity
constraints to the rank minimization problem (\ref{eq:7}). When $\bx$
is sparse, the result of the outer product $\bbx=\bx\bx^{*}$ is a
sparse matrix as well, as shown in Fig. \ref{fig:Sparse-vector-outer}.
Therefore, one strategy might be to minimize the $l_{1}$ norm of
the matrix $\bbx.$ Alternatively, it is possible to exploit further
the structure of $\bbx,$ by noticing that the number of rows in $\bbx$
with a nonzero norm is equal to the number of non-zero values in $\bx$.
This means that sparsity of $\mathbf{x}$ also implies a small number
of non-zero rows in $\bbx$. Consider the vector $\mathbf{p}$ containing
the $l_{2}$ norm of the rows of $\bbx$, i.e. $p_{j}=\big(\sum_{k}|X_{jk}|^{2}\big)^{\frac{1}{2}}$
(note that the $l_{2}$ norm can be replaced by any other norm). Since
$\mathbf{p}$ should be sparse, one might try to impose a low $l_{1}$
norm on $\mathbf{p}$, in the spirit of $l_{1}$ minimization for
the sparse linear problem. This yields the constraint $||\mathbf{p}||_{1}=\sum_{j}|p_{j}|=\sum_{j}\big(\sum_{k}|X_{jk}|^{2}\big)^{\frac{1}{2}}\leq\eta$,
corresponding exactly to a low mixed $l_{1-2}$ norm constraint on
$\bbx$ \cite{duarte2011structured}. The problem to solve, as cast
in \cite{shechtman_sparsity_2011-3} is therefore :

\begin{equation}
\begin{array}{ll}
\textnormal{\ensuremath{\min}} & \textnormal{rank}(\bbx)\\
\mbox{s.t.} & |\textnormal{Tr}(\bba_{k}\bbx)-y_{k}|\leq\epsilon,\,\,\, k=1,\dots,M,\\
 & \bbx\succeq0,\\
 & \sum_{j}\big(\sum_{k}|X_{jk}|^{2}\big)^{\frac{1}{2}}\leq\eta,
\end{array}\label{eq:QCS}
\end{equation}
where $\epsilon$ is a noise parameter, and $\eta$ is a sparsity
parameter, enforcing row sparsity of $\mathbf{X}$. 

Since finding a rank 1 matrix $\mathbf{X}$ satisfying the constraints
is NP hard, the solution to (\ref{eq:QCS}) is approximated in (\cite{shechtman_sparsity_2011-3})
using the iterative log-det heuristic proposed in \cite{fazel2003log},
with an additional thresholding step added at each iteration, to further
induce signal sparsity. Once a low rank matrix $\hat{\mathbf{X}}$
that is consistent with the measurements and the sparse priors is
found, the sought vector $\mathbf{x}$ is estimated by taking the
best rank 1 approximation of $\hat{\bbx}$ using the singular value
decomposition (SVD): Decomposing $\hat{\bbx}$ into $\hat{\bbx}=\mathbf{USV^{T}}$,
the rank-1 approximation of $\hat{\bbx}$ is taken as $\mathbf{\hat{X}_{1}}=S_{11}\mathbf{U_{1}}\mathbf{U_{1}}^{*}$,
where $S_{11}$ represents the largest singular value, and $ $$\mathbf{U_{1}}$
is the corresponding column of $\mathbf{U}$.

Similar ideas that add sparse priors to SDP methods have been later
suggested in \cite{ohlsson_compressive_2011,jaganathan_recovery_2012-1,waldspurger_phase_2012}.
In \cite{ohlsson_compressive_2011}, the rank minimization objective
is relaxed to a convex trace minimization, with an additional $l_{1}$
regularization term to induce sparsity. This formulation yields:

\begin{equation}
\begin{array}{ll}
\textnormal{min} & \textnormal{Tr}(\mathbf{X})+\lambda||\mathbf{X}||_{1}\\
\mbox{s.t.} & |\textnormal{Tr}(\bba_{k}\mathbf{X})-y_{k}|\leq\epsilon\,\,\, k=1,\dots,M,\\
 & \mathbf{X}\succeq0.\\
\\
\end{array}\label{eq:CPRL}
\end{equation}
The solution of (\ref{eq:CPRL}) is shown \cite{ohlsson_compressive_2011}
to be unique in the noiseless case ($\epsilon=0$), under the following
condition: $||\bar{\bbx}||_{0}\leq\frac{1}{2}\big(1+\frac{1}{\mu})$,
where $\bar{\bbx}=\bar{\bx}\bar{\bx}^{*}$, with $\bar{\bx}$ being
the true solution to (\ref{eq:4}). The mutual coherence $\mu$ is
defined by $\mu=\max_{i,j}\frac{<\mathbf{B}_{i},\mathbf{B}_{j}>}{||\mathbf{B}_{i}||||\mathbf{B}_{j}||}$
, with $\mathbf{B}$ being the matrix satisfying $\by=\mathbf{BX^{S}}$,
where $\mathbf{X^{S}}$ is the vector obtained from stacking the columns
of $\mathbf{X}$. The same work also relates other recovery guarantees
to the RIP criterion. 

In \cite{li2012sparse} it is shown that for $\ba_{i}$ that are independent,
zero-mean normal vectors, on the order of $k^{2}\log n$ measurements
are sufficient to recover a $k$-sparse input from measurements of
the form (\ref{eq:4}), using SDP relaxation. In \cite{jaganathan_recovery_2012-1},
an algorithm is suggested to solve the sparse 1D Fourier phase retrieval
problem based on a two-step process, each step cast separately as
an SDP problem: First, the support of $\bx$ is determined from its
autocorrelation sequence, and then $\bx$ is determined, given the
support. This algorithm is shown experimentally to recover $k$ sparse
signals from $O(k^{2})$ measurements.

\subsubsection*{Greedy methods with sparsity prior}

Since the matrix lifting algorithms involve a dimension increase,
they are not ideally suited for large vectors, where computational
cost can become significant. In addition, they are in general not
guaranteed to converge to a solution. An alternative to the SDP algorithms
is posed by sparsity based greedy algorithms \cite{bahmani_greedy_2011,beck_sparsity_2012-1,shechtman_gespar:_2013}.
One algorithm, that is both fast and accurate, is a greedy method
named GESPAR (for GrEedy Sparse PhAse Retrieval) \cite{shechtman_gespar:_2013}.
GESPAR attempts to solve the least squares sparse quadratic problem
(\ref{eq:LS}). Namely, it seeks a $k$ sparse vector $\mathbf{x}$
consistent with the quadratic measurements $\mathbf{y}$. GESPAR is
a fast local search method, based on iteratively updating signal support,
and seeking a vector that corresponds to the measurements, under the
current support constraint. A local-search method is repeatedly invoked,
beginning with an initial random support set. Then, at each iteration
a swap is performed between a support and an off-support index. Only
two elements are changed in the swap (one in the support and one in
the off-support), following the so-called \emph{2-opt} method \cite{papadimitriou1998combinatorial}.
Given the support of the signal, the phase-retrieval problem is then
treated as a non-convex optimization problem, approximated using the
damped Gauss Newton method \cite{bertsekas_nonlinear_1999-1}. See
Algorithm\textbf{ \ref{gespar}} for a detailed description of the
algorithm. 

\begin{figure*}
\centering{}\textbf{}%
\Ovalbox{\begin{minipage}[t]{1\textwidth}%
\begin{center}
\textbf{Sparse phase retrieval algorithms - a comparison}
\par\end{center}

\begin{singlespace}
We simulate sparse-Fienup \cite{mukherjee_iterative_2012} and GESPAR
\cite{shechtman_gespar:_2013} for various values of $N\in[64,2048]$,
and $M=2N$. The recovery probability vs. sparsity $k$ for different
vector lengths is shown in Figs.~\ref{fig:Sparsity-based-phase-retrieval}a
and ~\ref{fig:Sparsity-based-phase-retrieval}b. In both cases the
recovery probability increases with $N$, while GESPAR clearly outperforms
the alternating iteration method.

We then simulate the recovery success rate of three sparsity-based
phase retrieval algorithms. We choose $\bx$ as a random vector of
length $N=64$. The vector contains uniformly distributed values in
the range $[-4,-3]\cup[3,4]$ in $k$ randomly chosen elements. The
$M=128$ point DFT of the signal is calculated, and its magnitude-square
is taken as $\by$, the vector of measurements. In order to recover
the unknown vector $\bx$, three methods are used: A greedy method
(GESPAR\cite{shechtman_gespar:_2013}), an SDP based method (Algorithm
2, \cite{jaganathan_recovery_2012-1}), and an iterative Fienup algorithm
with a sparsity constraint (\cite{mukherjee_iterative_2012}). The
Sparse-Fienup algorithm is run using $100$ random initial points,
out of which the chosen solution is the one that best matches the
measurements. Namely, $\hat{\mathbf{x}}$ is selected as the $s$
sparse output of the Sparse-Fienup algorithm with the minimal cost
$f(\bx)=\sum_{i=1}^{N}(|\bbf_{i}\bx|^{2}-y_{i})^{2}$ out of the $100$
runs. The probability of successful recovery is plotted in Fig. \ref{fig:Sparsity-based-phase-retrieval}c
for different sparsity levels $k$. The success probability is defined
as the ratio of correctly recovered signals $\bx$ out of $100$ simulations.
In each simulation both the support and the signal values are randomly
selected. The three algorithms (GESPAR, SDP and Sparse-Fienup) are
compared. The results clearly show that GESPAR outperforms the other
methods in terms of probability of successful recovery - over 90\%
successful recovery up to $k=15$, vs. $k=8$ and $k=7$ in the other
two techniques.

For more extensive comparisons, the reader is referred to \cite{shechtman_gespar:_2013}.
\end{singlespace}

\begin{singlespace}
\begin{center}
\includegraphics[width=0.9\textwidth]{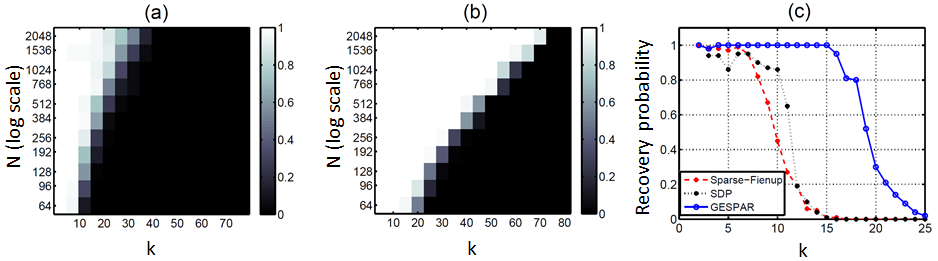}
\par\end{center}
\end{singlespace}

\begin{singlespace}
\caption{\label{fig:Sparsity-based-phase-retrieval}Sparsity-based phase retrieval
algorithms, a comparison. (a) Sparse-Fienup recovery probability vs.
sparsity $k$, for various signal length $N$, and with $M=2N$. (b)
GESPAR recovery probability vs. sparsity $k$, for various signal
length $N$, and with $M=2N$. (c) Recovery probability for three
algorithms: sparse-Fienup, SDP, and GESPAR for $N=64,\, M=128$ \cite{shechtman_gespar:_2013}.}

A major advantage of greedy methods over other algorithms (e.g. SDP
based) is its low computational cost; GESPAR may be used to find a
sparse solution to the 2D Fourier phase retrieval - or phase retrieval
of images. Figure \ref{fig:GESPAR_circles} shows a recovery example
of a sparse $195\times195$ pixel image, comprised of $s=15$ circles
at random locations and random values on a grid containing $225$
points, recovered from its $38,025$ 2D-Fourier magnitude measurements,
using GESPAR. The dictionary used in this example contains 225 elements
consisting of non-overlapping circles located on a $15\times15$ point
Cartesian grid, each with a 13 pixel diameter. The solution took 80
seconds. Solving the same problem using the sparse Fienup algorithm
did not yield a successful reconstruction, and using the SDP method
is not practical due to the large matrix sizes.
\end{singlespace}

\begin{singlespace}
\begin{center}
\includegraphics[width=0.5\textwidth]{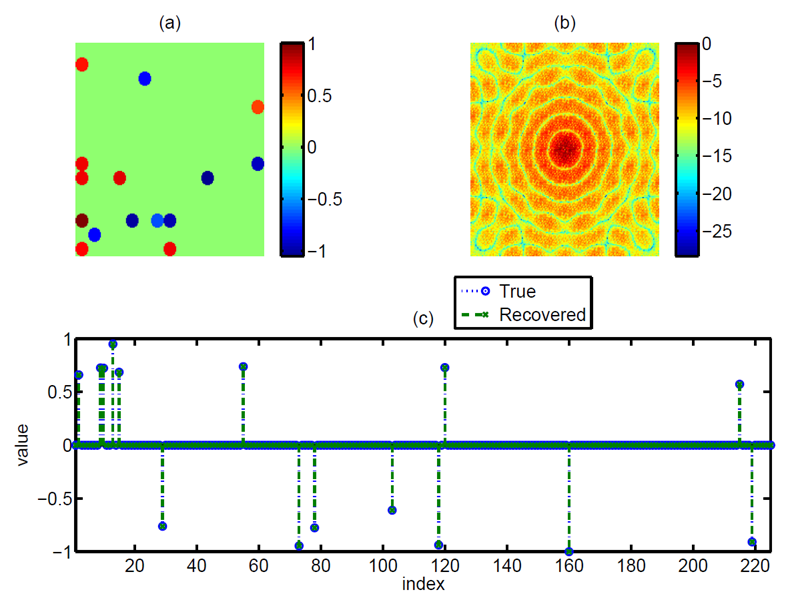}
\par\end{center}
\end{singlespace}

\begin{singlespace}
\caption{\label{fig:GESPAR_circles}2D Fourier phase retrieval example. (a)
True $195\times195$ sparse circle image ($s=15$ circles). (b) Measured
2D Fourier magnitude ($38,025$ measurements, log scale). (c) True
and recovered coefficient vectors, corresponding to circle amplitudes
at each of the $225$ grid points\cite{shechtman_gespar:_2013}.}
\end{singlespace}
\end{minipage}}
\end{figure*}

GESPAR has been shown to yield fast and accurate recovery results
(see \emph{Sparse phase retrieval algorithms} box and Fig. \ref{fig:Sparsity-based-phase-retrieval}
within\textbf{)}, and has been used in several phase-retrieval optics
applications - including CDI of 1D objects \cite{Sidorenko:13}, efficient
CDI of sparsely temporally varying objects \cite{shechtman_efficient_2013},
and phase retrieval via waveguide arrays \cite{Shechtman:13}. A similar
method has been used to solve the combined phase-retrieval and sub-wavelength
imaging problem \cite{szameit_sparsity-based_2012} (See sub-Section\emph{
}\ref{sub:Sub-wavelength-CDI-using}). 

\begin{algorithm}
\begin{singlespace}
\caption{\label{gespar}GESPAR - main steps}

\textbf{Input:} $\bba_{i},y_{i},\tau,{\rm ITER}$.\\
 $\bba_{i}\in\real^{N\times N},i=1,2,\ldots,M$ - symmetric matrices.\\
 $y_{i}\in\real,i=1,2,\ldots,M.$\\
 $\tau$ - threshold parameter. \\
 ITER - Maximum allowed total number of swaps. \\
\textbf{Output:} $\bx$ - an optimal (or suboptimal) solution of (\ref{eq:LS}).

\line(1,0){250}

\textbf{Initialization}. Set $T=0,j=0$. 
\end{singlespace}
\begin{enumerate}
\begin{singlespace}
\item Generate a random index set $S_{0}\,(|S_{0}|=s)$
\item Invoke the damped Gauss Newton method with support $S_{0}$, and obtain
an output $\bz_{0}$. Set $\bx_{0}=\bbu_{S_{0}}\bz_{0}$, where $\bbu_{S_{0}}\in\mathbb{R}^{N\times s}$
is the matrix consisting of the columns of the identity matrix $\bbi_{N}$
corresponding to the index set $S_{0}$\end{singlespace}

\end{enumerate}
\begin{singlespace}
\textbf{General Step ($j=1,2,\ldots$):}
\end{singlespace}
\begin{enumerate}
\begin{singlespace}
\item [3)]\setcounter{enumi}{3}Update support: Let $p$ be the index from
$S_{j-1}$ corresponding to the component of $\bx_{j-1}$ with the
smallest absolute value. Let $q$ be the index from $S_{j-1}^{c}$
corresponding to the component of $\nabla f(\bx_{j-1})$ with the
highest absolute value, where $\nabla f(\bx)$ is the gradient of
the least-squares objective function from (\ref{eq:LS}), namely $\nabla f(\bx)=4\sum_{i}(\mathbf{x}^{*}\mathbf{A}_{i}\mathbf{x}-y_{i})\mathbf{A}_{i}\mathbf{x}$
. Increase $T$ by 1, and make a swap between the indices $p$ and
$q$, i.e. set $\tilde{S}$ to be: 
\[
\tilde{S}=(S_{j-1}\backslash\{p\})\cup\{q\}.
\]

\item Minimize with given support: Invoke the damped Gauss Newton method
\cite{bertsekas_nonlinear_1999-1} with input $\tilde{S}$ and obtain
an output $\tilde{\bz}$. Set $\tilde{\bx}=\bbu_{S}\tilde{\bz}$,
where $\bbu_{S}\in\mathbb{R}^{N\times s}$ is the matrix consisting
of the columns of the identity matrix $\bbi_{N}$ corresponding to
the index set $S$. \\
 If $f(\tilde{\bx})<f(\bx_{j-1})$, then set $S_{k}=\tilde{S},\bx_{k}=\tilde{\bx}$,
advance $m$ and go to 3. If none of the swaps resulted with a better
objective function value, go to 1.\end{singlespace}

\end{enumerate}
\begin{singlespace}
\textbf{Until} $f(\bx)<\tau$ or $T>{\rm ITER}$. 

The output is the solution $\bx$ that yielded the minimum value for
the least-squares objective.\end{singlespace}
\end{algorithm}

\newpage{}

\clearpage

\section{Applications in Lensless Imaging }

In this section, we present several CDI applications with connection
to the phase retrieval algorithms described in previous sections.
The concept of phase retrieval in optical imaging arises from the
attempt to recover images from experimental measurements. To this
end, it is essential to emphasize that compared to numerical simulations
or signal processing of digital data, phase retrieval of experimentally
obtained patterns has several additional challenges. First, the far-field
intensity distribution (Fourier magnitude) is corrupted by various
types of noise, such as Poisson noise, detector read-out noise, and
unwanted parasitic scattering from the optics components in the system.
Second, in single-shot experiments, the measured far-field intensity
distribution is usually incomplete, including a missing center (i.e.
the very low spatial frequency information cannot be directly recorded
by a detector) \cite{miao2005quantitative}. Third, when the far-field
intensity distribution is measured by a detector, each pixel integrates
the total number of photons within the solid angle subtended by the
pixel, which is not exactly equivalent to uniform sampling of the
diffraction signal \cite{song2007phase}. Additionally, many experiments
are carried out using incoherent (but bright) sources. Coherence is
achieved by propagating a long distance from the source, but often
the experiment is constrained to be carried out with a partially-incoherent
beam \cite{whitehead2009diffractive}. All of these issues add complications
to algorithmic phase retrieval. However, notwithstanding these challenges,
successful phase retrieval of experimental data in optical imaging
has been widely achieved \cite{miao_extending_1999-1,robinson2001reconstruction,zuo2003atomic,williams2003three,chapman_femtosecond_2006,song_quantitative_2008-1,dierolf_ptychographic_2010,seibert_single_2011,loh_fractal_2012}.
Below, we show several examples.

\subsection{\label{sub:Quantitative-comparisons-of}Quantitative comparison of
alternating-projection algorithms }

\begin{figure}
\begin{centering}
\includegraphics[width=0.9\columnwidth]{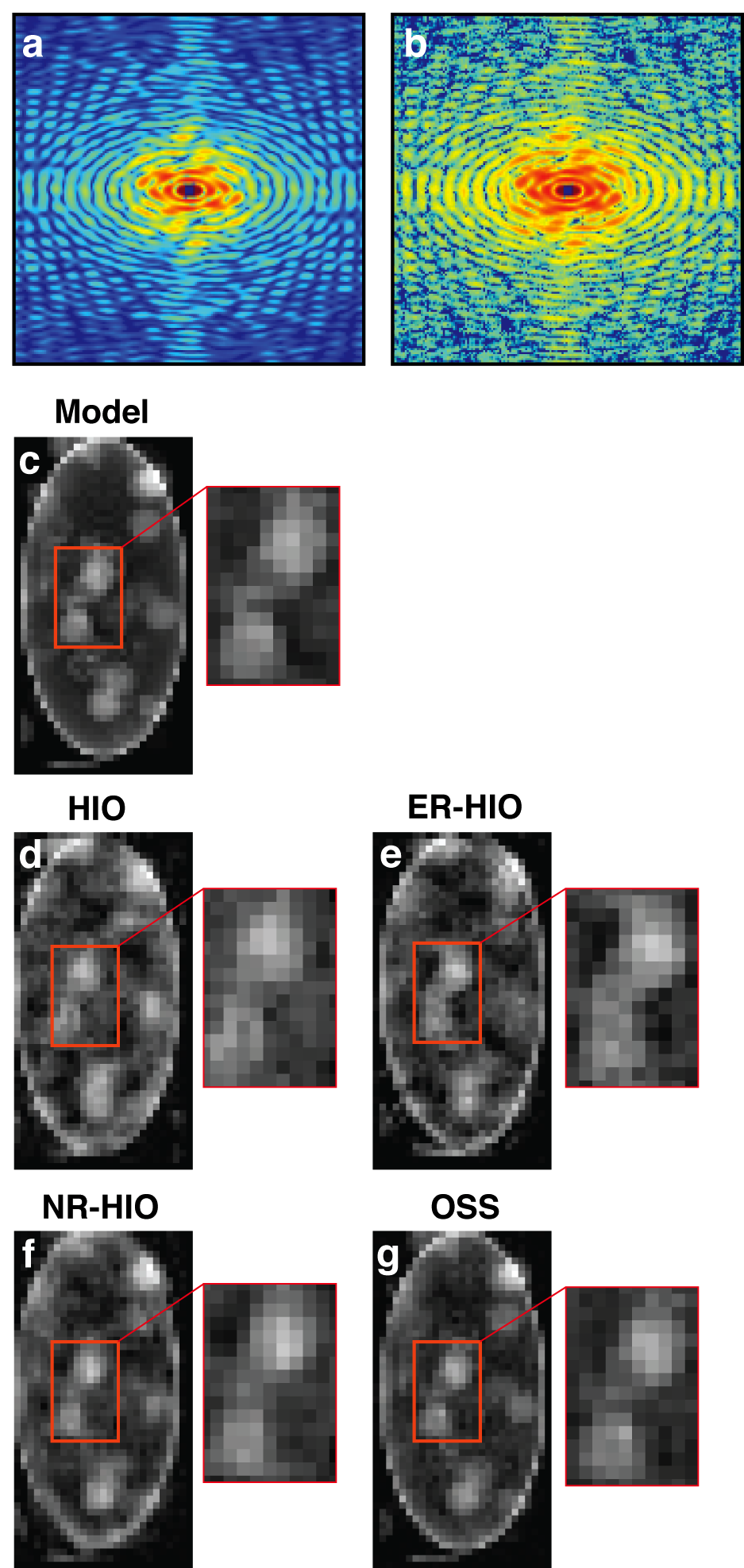}
\par\end{centering}

\caption{\label{fig:A-quantitative-comparison}A quantitative comparison between
the HIO, ER-HIO, NR-HIO, and OSS algorithms. (a) Noise-free oversampled
diffraction pattern calculated from simulated biological vesicle.
(b) High Poisson noise added to the oversampled diffraction pattern.
(c) The structure model of the biological vesicle and its fine features
(inset). The final reconstruction of the noisy diffraction pattern
(b) by (d) HIO, (e) ER-HIO, (f) NR-HIO, (g) and OSS \cite{rodriguez2013oversampling}.}

\end{figure}

Quantitative comparisons between the OSS, HIO, ER-HIO and NR-HIO algorithms
have been performed using both simulated and experimental data \cite{rodriguez2013oversampling}.
Figure \ref{fig:A-quantitative-comparison} shows a noise-free oversampled
diffraction pattern (Fourier magnitude squared) calculated from a
simulated biological vesicle (Fig. \ref{fig:A-quantitative-comparison}c).
High Poisson noise was then added to the diffraction intensity (Fig.
\ref{fig:A-quantitative-comparison}b). Figures \ref{fig:A-quantitative-comparison}d-g
show the final reconstructions by HIO, ER-HIO, NR-HIO, and OSS, respectively.
Visually, OSS produced the most faithful reconstruction among the
four algorithms (insets in Fig. \ref{fig:A-quantitative-comparison}d-g).
The recovery error was quantified using consistency with the measurements:
\begin{equation}
E=\sum_{n}|z_{r}[n]-z_{m}[n]|/\sum_{n}|z_{m}[n]|
\end{equation}
where $z_{r}[n]$ is the final reconstruction and $z_{m}[n]$ is the
model structure. The value for $E$ of the HIO, ER-HIO, NR-HIO, and
OSS reconstructions is $0.28,0.24,0.16$ and $0.07$, respectively.

\begin{figure*}

\centering{}\includegraphics[width=0.8\textwidth]{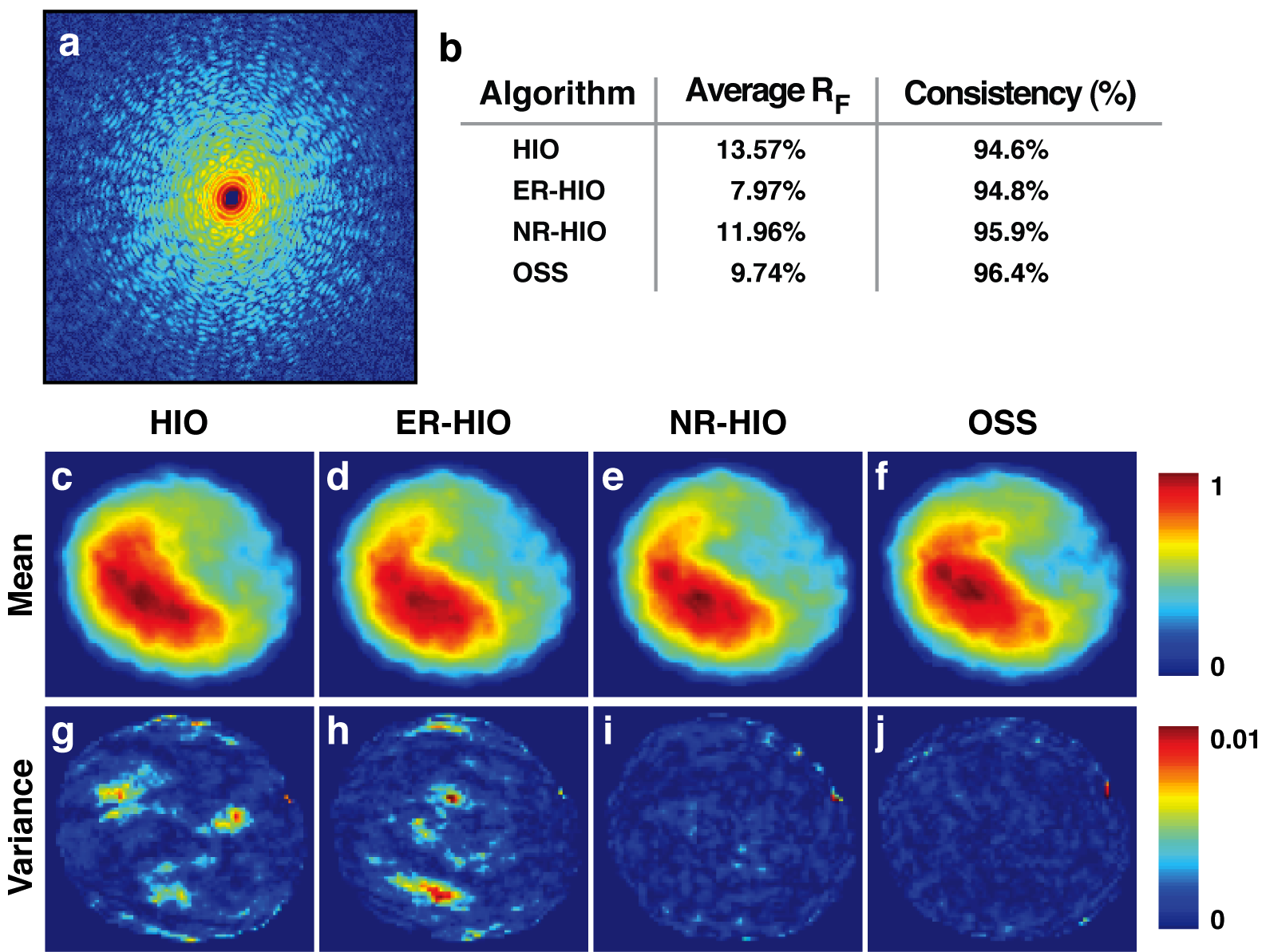}\caption{\label{fig:Exp_yeast}Phase retrieval of an experimental diffraction
pattern from a biological sample. (a) Oversampled X-ray diffraction
pattern measured from a S. pombe yeast spore cell. (b) The average
$R_{F}$ and the consistency of five independent trials of phase retrieval
using four different algorithms. The average reconstruction of five
independent trials using HIO (c), ER-HIO (d), NR-HOP (e), and OSS
(f). The variance of five final images with HIO (g), ER-HIO (h), NR-HOP
(i), and OSS (j). \cite{rodriguez2013oversampling} }
\end{figure*}

Next, the four algorithms were compared using an experimental diffraction
pattern measured from a Schizosaccharomyces pombe yeast spore cell
\cite{rodriguez2013oversampling}. The experiment was conducted on
an undulator beamline at a 3rd generation synchrotron radiation (Spring-8)
in Japan. A coherent wave of 5 keV X-rays was incident on a fixed,
unstained S. pombe yeast spore. An oversampled X-ray diffraction pattern
was acquired by a charge-coupled device detector. Figure \ref{fig:Exp_yeast}a
shows the experimental diffraction pattern in which the centro-square
represents the missing low spatial resolution data \cite{jiang2010quantitative}.
By using a loose support, phase retrieval was performed on the measured
data with the HIO, ER-HIO, NR-HIO, and OSS algorithms. For each algorithm,
five independent trials were conducted, each consisting of 100 independent
runs with different random initial phase sets. In each trial, the
reconstruction with the smallest error metric $R_{F}$ was chosen
as a final image, where $R_{F}$ is defined as:

\begin{equation}
R_{F}=\sum_{k}\bigg||Z_{e}[k]|-\zeta|Z_{r}[k]|\bigg|/\sum_{k}\big|Z_{e}[k]\big|.
\end{equation}
Here $|Z_{e}[k]|$ is the measured Fourier magnitude, $|Z_{m}[k]|$
is the recovered Fourier magnitude, and $\zeta$ is a scaling factor. 

For each algorithm, the mean and average of the five final images
were used to quantify the reconstruction. Figures \ref{fig:Exp_yeast}c-j
show the average and variance of five final images obtained by HIO,
ER-HIO \cite{fienup_phase_1982}, NR-HIO \cite{Martin:12}, and OSS
\cite{rodriguez2013oversampling}, respectively. The average $R_{F}$
and the consistency of five independent trials are shown in Fig. \ref{fig:Exp_yeast}b.
Both visual inspection and quantitative results indicate that OSS
produced the most consistent reconstructions among all four algorithms. 

\subsection{X-ray free electron laser CDI}

The majority of imaging experiments at X-ray free-electron laser sources
utilize the method of CDI. The lensless nature is actually an advantage
when dealing with extremely intense and destructive pulses, where
one can only carry out a single pulse measurement with each object
(say, a molecule) before the object disintegrates. In such cases,
often one cannot use any optical components at all, because any component
(e.g., a lens) would be severely damaged by the extremely high flux
of (X-ray) photons. CDI solves these problems: it works without the
need for optical components. In this vein, CDI also facilitates reliable
imaging of moving objects. Indeed, in many experiments the objects
move (flow) across the X-ray beam, for example, when the X-ray laser
beam hits a focused aerosol beam or nano-particles in a liquid jet.
In such an experiment, the particle density is usually adjusted so
that the X-ray laser pulse is more likely to hit a single particle
than several. A particle is hit by chance by a pulse, but this is
not known until the diffraction pattern is read out from the detector,
which is done on every pulse. The stream of data is then analyzed
and sorted to give the single-particle hits, which contain the meaningful
measured data, while all other data is ignored.

There are two generic classes of these \textquotedblleft{}single particle\textquotedblright{}
CDI experiments: imaging of reproducible particles, and imaging of
unique particles. The first category includes particles such as viruses.
Assuming that these particles are not aligned in the same direction,
the collected data represents diffraction patterns of a common object,
but in random orientations. If the orientations can be determined
then the full three-dimensional Fourier magnitude of the object can
be determined, which in turn could be phased to give a 3D image. A
proof of concept of this experiment was carried out by Loh et al \cite{loh2010cryptotomography}.

\begin{figure*}
\begin{centering}
\includegraphics[width=0.8\textwidth]{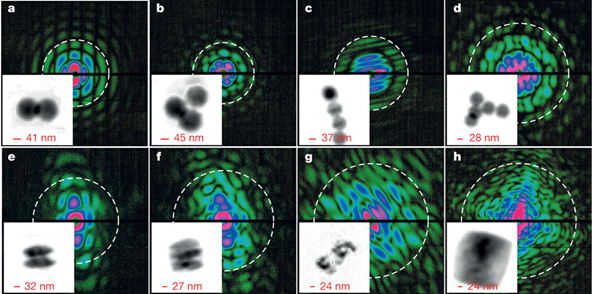}
\par\end{centering}

\caption{\label{fig:Diffraction-patterns-of}Diffraction patterns from single
X-ray FEL pulses from particles in flight, and reconstructed images.
a\textendash{}d, Clusters of polystyrene spheres with radii of 70
nm (a, b) and 44 nm (c, d). e, f, Ellipsoidal nanoparticles. g, A
soot particle. h, A salt\textendash{}soot mixture \cite{loh_fractal_2012}.}

\end{figure*}

An example of the second class of \emph{flash diffractive imaging}
is imaging airborne soot particles in flight in an aerosol beam \cite{loh_fractal_2012}.
Several diffraction patterns of soot particles and clusters of polystyrene
spheres (as test objects) are shown in Fig. \ref{fig:Diffraction-patterns-of},
along with the 2D reconstructions of the objects. The experiments
were carried out at the LCLS, using the CFEL-ASG Multi-Purpose (CAMP)
instrument \cite{struder2010large} at the Atomic, Molecular and Optical
Science beam line \cite{bozek2009amo}. Pulses of about $10^{12}$
photons of 1.0 nm wavelength were focused to an area of 10 $\textnormal{\textnormal{\ensuremath{\mu}}\ensuremath{m^{2}}}$.
The X-ray detectors (pnCCD panels) were placed to give a maximum full-period
resolution of 13 nm at their center edges. 

In these experiments, the phase retrieval of the patterns was carried
out using the Relaxed Averaged Alternating Reflections (RAAR) \cite{luke2005relaxed}
algorithm, and using the Shrinkwrap procedure  \cite{PhysRevB.68.140101},
which determines and iteratively updates the support constraint used.
The objects were such that it was possible to apply an additional
constraint that the image is real valued. Strikingly, the X-ray coherent
diffraction patterns have very high contrast. The intensity minima
are close to zero. This has an enormous effect on the ability to recover
the phase of these pattern reliably. This reliability is quantified
in the phase-retrieval transfer function (PRTF) \cite{chapman2006high},
which compares the magnitude of the complex-valued average of patterns
phased with different starting guesses to the square root of the measured
diffraction pattern. If, at a particular pixel of the diffraction
pattern, the phases are consistently reconstructed, then the sum over
$N$ patterns will give a magnitude $N$ times higher than the measured
magnitude, and so the PRTF will be unity. If the phases are random,
then this sum will tend to zero. For patterns generated with X-Ray
free electron lasers, this function is often close to unity and is
lower primarily in areas where the signal to noise is low. This limited
signal is what ultimately limits the resolution; an estimate of the
achieved resolution is given by the white dotted circle on each pattern
in Fig. \ref{fig:Diffraction-patterns-of}. The reconstructed images
are sums of ten independent reconstructions. These complex-valued
sums have the nice property that the their Fourier spectrum is effectively
modulated by the PRTF and hence any artifacts due to noise (or even
due to forcibly truncating the data to a lower resolution) is unlikely
to show up in the recovered image. 

\subsection{Tabletop short wavelength CDI}

\begin{figure*}
\begin{centering}
\includegraphics[width=0.8\textwidth]{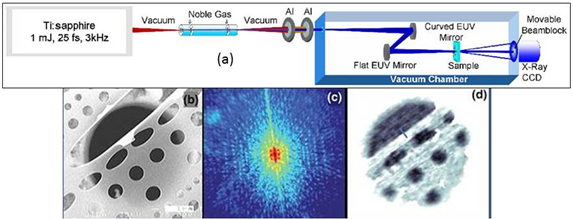}
\par\end{centering}

\caption{\label{fig:First-tabletop-short-wavelength}First tabletop short-wavelength
coherent diffraction imaging. (a) Experimental setup. Coherent extreme
UV radiation is generated through the process of high harmonic generation.
A single harmonic order at wavelength 29 nm is selected and focused
onto a sample by a pair of multilayer mirrors. The scattered light
is detected by X-ray CCD camera. (b) The original image, used to analyze
the performance of the CDI process, obtained with a Scanning electron
Microscope (SEM). The image shows a masked carbon film placed on a
15 \textmu{}m diameter pinhole. (c) Recorded multi-frame diffraction
pattern (corresponding to Fourier magnitude squared of the object
shown in (b)). (d) CDI reconstruction using the HIO algorithm with
214 nm resolution \cite{sandberg_lensless_2007}.}

\end{figure*}

To-date, most CDI experiments are carried out in 3rd generation synchrotron
and X-ray free electron lasers. However, limited access and experimental
time hinder the development and applications of CDI using these methods.
Thus, over the past several years, CDI microscopes that are based
on tabletop sources of coherent extreme UV and soft X-rays are also
being developed \cite{sandberg2013studies}. Figure \ref{fig:First-tabletop-short-wavelength}
shows the first tabletop CDI experiment with extreme UV wavelength. 

\subsection{Sub-wavelength CDI using sparsity\label{sub:Sub-wavelength-CDI-using}}

Prior knowledge of object-sparsity can help regularize the phase-retrieval
problem, as well as compensate for loss of other kinds of information
- in this example - the loss of high spatial frequencies. As described
before, when an object is illuminated by coherent light of wavelength
$\lambda$, the far-field intensity pattern is proportional to the
magnitude of the object's Fourier transform. In addition, features
in the object that are smaller than $\sim\lambda/2$ are smeared due
to the diffraction limit. Consequently, the intensity measured in
the far field corresponds to $\mathbf{y}\propto|\mathbf{LF}\mathbf{x}|^{2}$
where $\mathbf{L}$ represents a low pass filter at cutoff frequency
$\nu_{c}=1/\lambda$, $\mathbf{F}$ represents the Fourier transform,
and $|\cdot|^{2}$ stands for elementwise squared absolute value. 

\begin{figure*}
\begin{centering}
\includegraphics[width=0.8\textwidth]{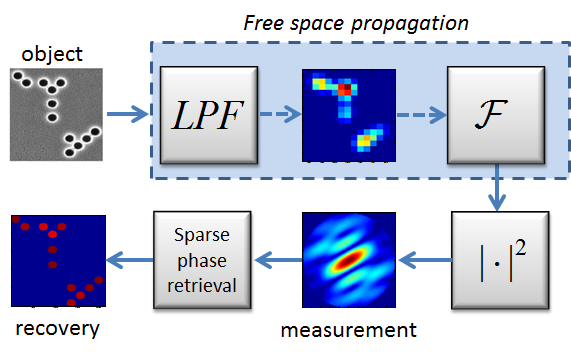}
\par\end{centering}

\caption{\label{fig:Sparsity-based-sub-wavelength}Sparsity based sub-wavelength
CDI. A 2D object consisting of an arrangement of nano-holes (100nm
in diameter) is illuminated by a 532nm laser, and the Fourier plane
magnitude is measured. High spatial frequencies are lost during propagation,
because the features (the circles as well as their separation) are
smaller than $\sim\lambda/2$. Using an iterative greedy algorithm,
and exploiting the prior knowledge that the object is sparse in a
dictionary made of 100nm circles, the phase is retrieved and the object
is recovered from its low-pass-filtered Fourier magnitude \cite{szameit_sparsity-based_2012}. }
\end{figure*}

Figure \ref{fig:Sparsity-based-sub-wavelength} (adapted from \cite{szameit_sparsity-based_2012}),
shows the recovery of a sparse object containing sub-wavelength features
($100nm$ holes illuminated by a $\lambda=532nm$ laser) from its
experimentally measured low pass filtered Fourier magnitude. The prior
knowledge used for recovery is that the object is comprised of a small
number of $100nm$ diameter circles on a grid, illuminated by a plane
wave. The exact number, locations, and amplitudes of the circles are
not known a-priori. The recovery is performed using a greedy algorithm
that iteratively updates the support of the object, finds a local
minimum and removes the weakest circle, until convergence \cite{szameit_sparsity-based_2012}.

Another type of information loss in CDI, for which the prior knowledge
of object sparsity can be helpful, is low signal to noise ratio. In
non-destructive X-ray CDI measurements, it is not uncommon for signal
acquisition time to be on the order of tens of seconds \cite{sandberg_lensless_2007,stadler_hard_2008,seaberg_ultrahigh_2011},
in order to acheive sufficiently high SNR. This poses a severe limitation
on the temporal resolution attainable with such measurements, restricting
the types of dynamical phenomena accessible by X-ray CDI. Expoilting
sparsity in the \emph{change }that an object undergoes between subsequent
CDI frames has been recently suggested as a means to overcome high
noise values, and consequently significantly decrease acquisition
time \cite{shechtman_efficient_2013}. In other words, the fact that
an object is \emph{sparsely varying}, can be used as prior information
to effectively denoise sequential Fourier magnitude measurements.
In \cite{shechtman_efficient_2013}, CDI of a sparsely varying object
is formulated as a sparse quadratic optimization problem, and solved
using GESPAR \cite{shechtman_gespar:_2013}. Numerical simulations
suggest a dramatic potential improvement in temporal resolution: In
an example consisting of a $51\times51$ pixel object with 5 randomly
varying pixels between frames, an improvement of 2 orders of magnitude
in acquisition time is possible \cite{shechtman_efficient_2013}.

\section{Other Physical Settings, bottlenecks, and Vision }

This review article focused mainly on the simplest physical setting
for phase retrieval in optical imaging (Fig. \ref{fig:CDI-setup:-A}),
CDI: where an unknown 2D optical image is recovered algorithmically
from a single measurement of its far-field intensity pattern, given
a known image support (or other prior information). In terms of signal
processing, this problem corresponds to recovering a 2D object from
measurements of its Fourier magnitude. However, the issue of phase
retrieval in optical imaging, and in a more general sense \textendash{}
in optics, is far broader, and includes other physical settings which
naturally translate into signal processing problems different than
the standard phase retrieval problem. This section provides a short
overview of those physical settings, defines the various problems
in terms of signal processing, and provides some key references. We
conclude with a discussion on the main challenges and bottlenecks
of phase retrieval in optical imaging, followed by an outlook for
the upcoming years and long term vision.

\subsection{Non-Fourier Measurements }

The simplest optical phase retrieval problem assumes that the measured
data corresponds to the Fourier magnitude. In optical settings, this
means that the measurements are taken in the Fourier domain of the
sought image, which physically means performing the measurements at
a plane sufficiently far away from the image plane (the so-called
\emph{far-field} or the \emph{Fraunhofer regime}), or at the focal
plane of an ideal lens \cite{saleh_fundamentals_2007}. In reality,
however, the measurements can be taken at any plane between the image
plane and the far field, which would yield intensity patterns that
are very different than the Fourier magnitude of the sought signal.
This of course implies that new (or revised) algorithms - beyond those
described in previous sections - must be used, which naturally raises
issues of conditions for uniqueness and convergence. At the same time,
these measurements have some interesting advantages, which can be
used wisely to improve the performance of phase retrieval. Let us
begin by describing the relevant physical settings. 

As stated earlier in this paper, the optical Fourier plane corresponds
to a plane sufficiently far away from where the object (the sought
signal) is positioned. \emph{Far away} here means asymptotically at
infinite distance from the object plane, or at the focal plane of
a lens. However, the entire propagation-evolution of electromagnetic
waves from any plane to any other plane (not only from the near field
to the far field) is known: it is fully described by Maxwell\textquoteright{}s
equations. As such, one can formulate the problem through a proper
transfer function (of the electromagnetic wave) that is different
than the Fourier transform. In this context, the most well studied
case is the regime of Fresnel diffraction, where the transfer function
is expressed in an integral form known as Fresnel integral \cite{saleh_fundamentals_2007}.
This regime occurs naturally at a range of distances away from the
object plane which naturally includes also the Fraunhofer regime where
the transfer function reduces to a simple Fourier transform. Going
beyond the Fresnel regime is also possible. This means that the (magnitude
squared of the) electromagnetic wave will be measured at some arbitrary
plane away from the object. A more general case arisees by expressing
the scalar transfer function of the light in a homogeneous medium,
at any plane $z$ as:

\begin{equation}
T(k_{x},k_{y},z)=\exp\big[-iz\sqrt{k^{2}-(k_{x}^{2}+k_{y}^{2})}\big]\label{eq:Transfer fun free space}
\end{equation}
Here $k=\omega/c$, with $\omega$ being the frequency of the light,
$c$ being the speed of light in the medium, and $k_{x},k_{y}$ describe
the transverse wavenumbers. The field at any arbitrary plane $z$,
$E(x,y,z)$, is then given by inverse Fourier transforming the spectral
function at that plane $F(k_{x},k_{y},z)$, which is related to the
spectrum at the initial plane by:
\[
F(k_{x},k_{y},z)=F(k_{x},k_{y},z=0)T(k_{x},k_{y},z).
\]

With the transfer function \ref{eq:Transfer fun free space}, one
can now formulate a new phase retrieval problem, where the measurements
are conducted at some arbitrary plane $z$, giving $|E(x,y,z)|^{2}$,
and the sought signal is $E(x,y,z=0)$. This approach can be extended
to include polarization effects, where the transfer-function is vectorial,
thereby describing the propagation through Maxwell\textquoteright{}s
equations with no approximation at all. The optical far-field - where
the measurement corresponds to the Fourier magnitude of the image
at the initial plane, (i.e., the measurement is proportional to $|F(k_{x},k_{y},z)|^{2}$)
- is obtained for distances $z$ larger than some minimum distance
$z_{0}$ that depends on the spectral extent of $F(k_{x},k_{y},z=0)$,
and only within a region close enough to the $z$ axis in the measurement
plane. 

It is interesting to compare these more general phase retrieval problems
to the generic problem of recovering a signal from its Fourier magnitude.
In terms of algorithmics, the generic problem is much simpler and
was extensively studied throughout the years, whereas the general
case is considerably more complex and was studied only sporadically.
However, in terms of optics, the measurements in the general case
can provide more information. Namely, in the general case, measurements
of $|E(x,y,z)|^{2}$ can be taken at multiple planes (multiple values
of $z$), and each measurement adds more information on the signal.
In contrast, for the generic problem, once the measurements are taken
in the optical far-field, taking more measurements at further away
distances does not add additional information because all the far
field measurements correspond to the Fourier magnitude (to within
some known scaling of coordinates in the measurement planes). As such,
performing phase retrieval of optical images in the most general (non-Fourier)
case can be beneficial, as it leads to multiple measurements, possibly
relaxing the conditions on oversampling and/or the advance knowledge
on the support in the image plane.

Historically, these ideas on non-Fourier measurements are known to
the optics community since the early days of optical phase retrieval
\cite{fienup_reconstruction_1978}. They are currently being used
in the context of improving the convergence of phase retrieval by
taking non-Fourier measurements at several planes \cite{williams2006fresnel,abbey2008keyhole}.
Alternatively, one can take measurements at several different optical
frequencies $\omega$, which would be expressed as different values
of $k=\omega/c$ in the general transfer function given above. In
this mutli-frequency context, it is important that the frequencies
are well separated, each having a narrow bandwidth, to conform the
high degree of coherence required for CDI. These ideas are now being
pursued by several groups \cite{whitehead2009diffractive,chen2009multiple,abbey2011lensless}.
Interestingly, the multi-frequency idea also works in the continuous
case of broad bandwidth pulses centered on a single frequency. In
this case, the power spectrum of the pulse must be known in advance
\cite{whitehead2009diffractive,abbey2011lensless,witte2013ultra}.
In a similar vein, recent work has demonstrated scanning CDI, where
the beam is scanned through overlapping regions on the sample to allow
imaging of extended objects, a method known as \emph{ptychography}
\cite{chapman1996phase,thibault2008high,dierolf_ptychographic_2010,peterson2012nanoscale}.

More sophisticated physical settings also exist, where the medium
within which the waves are propagating is not homogeneous in space.
Famous examples are photonic crystals, wherein the refractive index
varies periodically in space, in a known fashion, in one, two or three
dimensions. Obviously, in such settings the transfer function for
electromagnetic waves is fundamentally different from the transfer
function in free space. The phase retrieval problem in such systems,
albeit less commonly known, is no less important. For example, photonic
crystal fibers can in principle be used for imaging in endoscopy.
The measurements in such systems correspond to the magnitude squared
of the field at the measurement plane, which would be very different
than the Fourier magnitude of the sought image. Still, once the transfer
function is known, complicated as it may be, the phase retrieval problem
is well defined and can be solved with some modifications to the algorithms
described above. See, for example, pioneering work on phase retrieval
in a photonic crystal fiber \cite{shapira2005complete}, and very
recently on sparsity-based phase retrieval and super-resolution in
optical waveguide arrays \cite{Shechtman:13}. In addition to these,
the concept of CDI has also been extended to other schemes, such as
Bragg CDI, suitable to periodic images, to reconstruct the structure
and strain of nano-crystals \cite{robinson2009coherent,newton2009three,yang_coherent_2013,doi:10.1146/annurev-matsci-071312-121654}.

\subsection{Phase Retrieval Combining Holographic Methods }

As explained in the introduction, optical settings always suffer from
the inability of photodetectors to directly measure the phase of an
electromagnetic wave at frequencies of THz (terahertz) and higher.
Partial solution for this problem is provided through holography,
invented by Denis Gabor in 1948 \cite{gabor1948new} and awarded the
Nobel Prize in Physics in 1971. Holography involves interfering an
electromagnetic field carrying some image, $E_{image}$, with another
electromagnetic field of the same frequency and a known structure,
denoted as $E_{ref}$. Typically, the so-called reference wave, $E_{ref}$,
has a very simply structure, for example, approximately a plane wave
(wave of constant amplitude and phase). The detection system records
$|E_{image}+E_{ref}|^{2}$. Originally, such holographic recording
was done on a photographic plate which is made from a photosensitive
material whose transmission becomes proportional to the recorded pattern
$|E_{image}+E_{ref}|^{2}$. This photographic plate is called a \emph{hologram},
wherein the information contained in the image wave $E_{image}$ is
embedded in transmission function of the hologram. To see the recording,
the wave of the known pattern, $E_{ref}$, is generated (which is
possible because its structure is simple and fully known) and made
to illuminate the hologram. The magnitude of the wave transmitted
through the illuminated hologram is therefore proportional to $|E_{image}+E_{ref}|^{2}\cdot E_{ref}$.
One of the terms is therefore $|E_{ref}|^{2}\cdot E_{image}$. Since
$|E_{ref}|^{2}$ carries virtually no information (i.e., it is just
a constant), this transmitted wave reconstructs the image times that
constant. This is the principle of operation of holography. Over the
years, it has been shown that it is almost always beneficial to record
not the actual image but its Fourier spectrum, hence the reconstructed
information is the Fourier transform of the image, and the image itself
is recovered either in the far-field (as explained in the introduction)
or at the focal plane of a lens. This process is termed \emph{Fourier
holography} \cite{mcnulty1992high}. 

In the context of phase retrieval, holography is used for the purpose
of adding information in the measurement scheme. Because in most cases
the measurements used are Fourier magnitudes, which physically implies
far-field measurements, the natural inclusion of holographic methods
is through Fourier holography. For example, adding a tiny hole (a
delta function) at a predetermined position in the sample (close to
where the sought image resides) creates an additional wave in the
far field, with a tilted phase (arising from the displacement between
the hole and the sought image). The far field intensity therefore
now corresponds to the absolute value squared of the sum of the Fourier
spectrum of the sought image and the (known) wave. As such, it introduces
additional prior knowledge, which can be used for increased resolution
of the algorithmic recovery or for relaxing the constraints on the
prior knowledge on the support. These ideas have been exploited successfully
using X-rays and electrons by several groups \cite{Kikuta197286,eisebitt_lensless_2004,latychevskaia2012holography}.

\subsection{Challenges, Bottlenecks and Vision}

The current challenges can be briefly defined in a single sentence:
higher resolution, ability to recover more complex objects, improved
robustness to noise, and real-time operation. The very reason phase
retrieval in optical imaging has become so important is owing to the
vision to be able one day to image complex biological molecules directly,
track their structural evolution as it evolves in time, and even view
the dynamics of the electronic wave functions bonding atoms together.
The reasoning is obvious: to understand biology at the molecular level,
to decipher the secrets of how their atomic constituents bond together
and how they interact with other molecules. The current state of the
art is far from those goals: imaging resolution is not yet at the
atomic (sub-nanometer) level, and - at nanometric resolution - imaging
cannot handle objects that are bounded by a support that is extremely
large compared to the resolution. In terms of being able to perform
real-time experiments, state of the art measurements have demonstrated
extremely short optical pulses: tens of attoseconds ($10^{-18}$ seconds
- on the order of the passage of a photon through a distance comparable
to the size of an atom). Pioneering experiments have even started
to probe the dynamics of electrons in molecules and tunneling processes
on these time scales. But, as of today, none of these ultrafast methods
was applied to imaging of even a simple molecule, let alone complex
biological structures. 

Clearly, the underlying physics and engineering pose great challenges
to meet these goals. Generating coherent radiation in the hard X-ray
regime is still a major obstacle, often requiring very large enterprises
such as the X-ray sources at the SLAC National Accelerator Laboratory.
These facilities around the world are continuously improving their
photon flux at shorter wavelengths, thereby constantly improving imaging
resolution. The fundamental limits on the coherent X-ray flux possible
with current methods (such as synchrotrons, XFELs \cite{emma2010first,ishikawa2012compact},
and the process of high harmonics generation \cite{popmintchev2012bright})
are not even known. But the steady improvement does give hope for
imaging at the atomic level in the near future. Taking the CDI techniques
to the regime of attosecond science is an important challenge. These
pulses are extremely short, hence their bandwidth is huge, so the
coherent diffraction pattern is a superposition of their Fourier contents,
which requires new algorithmic methods. As described above, these
issues are currently being explored by several groups. But the problem
is fundamentally more complicated, because the process of scattering
of light by molecules at these short wavelengths and ultrashort timescales
is not like passing light through a mask on which an image is imprinted.
Rather, many issues related to light-matter interactions under these
conditions are yet to be understood (e.g., tunneling ionization of
atoms by laser pulses). 

Finally, the long term vision must include imaging the dynamics within
complex biological systems at the atomic level and in real time. But
such systems are extremely complex to handle, in terms of details
on many spatial and temporal scales simultaneously, in terms of the
statistical nature and huge redundancy in the physical processes taking
place within such complexes simultaneously, and even in terms of the
quantum mechanics governing the dynamics at those scales. This is
where the signal processing community can make a large impact, by
devising new and original methods for recovering the information from
experimental measurements. Clearly, the algorithms will have to be
tailored to the specific physical settings. 

\bibliographystyle{IEEEtran}
\bibliography{phaseRetReview}

\end{document}